\documentclass[acmlarge]{acmart}

\AtBeginDocument{%
  \providecommand\BibTeX{{%
    \normalfont B\kern-0.5em{\scshape i\kern-0.25em b}\kern-0.8em\TeX}}}

\setcopyright{acmcopyright}
\copyrightyear{2021}
\acmYear{2021}
\acmDOI{10.1145/1122445.1122456}

\usepackage{setspace}
\usepackage{appendix}
\usepackage{makecell}
\usepackage{rotfloat}
\usepackage{longtable}
\usepackage{pdflscape}
\usepackage{makecell}
\usepackage{color}
\usepackage{enumitem}
\setlist[itemize]{leftmargin=*}

\newcommand{\myet}{\textit{et al.}}
\newcommand{\myeg}{\textit{e.g.}}
\begin{document}

\title{Computational Technologies for Fashion Recommendation: A Survey}
\thanks{This work was supported in part by the Natural Science Foundation of China under Grant 61976145 and Grant 62272319. It is also supported by NExT++. P.~Y.~Mok is supported in part by the Research Grants Council of the Hong Kong SAR under Grant 152112/19E and by the Innovation and Technology Commission of Hong Kong, under grant ITP/028/21TP}
\author{Yujuan Ding}
\email{dingyujuan385@gmail.com}
\affiliation{
    \department{Department of Computing}
  \institution{The Hong Kong Polytechnic University}
  \city{Hong Kong SAR, China}}
  
\author{Zhihui Lai}
\authornote{Corresponding Author}
\email{lai_zhi_hui@163.com}
\affiliation{
 \department{College of Computer Science and Software Engineering}
  \institution{Shenzhen University}}
  \affiliation{
  \institution{Shenzhen Institute of Artificial Intelligence and Robotics for Society}
  \city{Shenzhen, China}}

\author{P.Y. Mok}
\email{tracy.mok@polyu.edu.hk}
\affiliation{
\department{School of Fashion and Textiles}
  \institution{The Hong Kong Polytechnic University}
  \city{Hong Kong SAR, China}}

\author{Tat-Seng Chua}
\email{dcscts@nus.edu.sg}
\affiliation{
\department{School of Computing}
  \institution{National University of Singapore}
  \city{Singapore}}

\begin{abstract}
Fashion recommendation is a key research field in computational fashion research and has attracted considerable interest in the computer vision, multimedia, and information retrieval communities in recent years. Due to the great demand for applications, various fashion recommendation tasks, such as personalized fashion product recommendation, complementary (mix-and-match) recommendation, and outfit recommendation, have been posed and explored in the literature. The continuing research attention and advances impel us to look back and in-depth into the field for a better understanding. In this paper, we comprehensively review recent research efforts on fashion recommendation from a technological perspective. We first introduce fashion recommendation at a macro level and analyse its characteristics and differences with general recommendation tasks. We then clearly categorize different fashion recommendation efforts into several sub-tasks and focus on each sub-task in terms of its problem formulation, research focus, state-of-the-art methods, and limitations. We also summarize the datasets proposed in the literature for use in fashion recommendation studies to give readers a brief illustration. Finally, we discuss several promising directions for future research in this field. Overall, this survey systematically reviews the development of fashion recommendation research. It also discusses the current limitations and gaps between academic research and the real needs of the fashion industry. In the process, we offer a deep insight into how the fashion industry could benefit from the computational technologies of fashion recommendation.
\end{abstract}

\maketitle

\section{Introduction}
\label{sec:introduction}
Fashion was considered important in France as early as the 15th century and has swept the globe ever since~\cite{svendsen2006fashion}. Over the past centuries, industrialization, globalization, democratization, and the development of technology have increased the demand for fashion, and have naturally invigorated the fashion industry~\cite{mcintyre2016multi}. Fashion today is an essential aspect of culture and life, and the fashion industry is one of the most important parts of the global economy. More recently, the advent and development of the Internet have provoked a storm of digital transformation in all segments of the economy, including fashion. Nowadays, consumers increasingly shop for apparel or footwear online as they seek greater convenience, novelty, and personalization. Such digital transformation produces a huge demand for supporting technologies such as fashion recommendation. It also boosts the development of related research fields~\cite{gu2020fashion, ding2022leveraging}. 
 
Fashion recommendation refers to the task of providing suitable suggestions to facilitate the decision-making process related to fashion, which can benefit both the consumers and the fashion business. For example, in online fashion shopping, consumers want to easily find the items they like and experience personalized services, which need the help of recommendation tools considering the large scale of fashion items available.  In addition, consumers lacking a sense of fashion may need appropriate dressing or matching advice, which is another form of fashion recommendation. From the business perspective, effective recommendation tools help fashion retailers better understand the styling preference of their customers so that they can suggest the right products or outfits for them at the right time, which can optimize the conversion, Average Order Value (AOV) and repeat purchase rate in the end. Fashion recommendation has already become prevalent in modern life as a large percentage of people rely upon advice from their friends, families, or sales staff in fashion shopping~\cite{liew2011socially,guan2016apparel}. Such a decision-making pattern demonstrates the importance of an effective recommender system in fashion, which acts as a shopping assistant to offer personalized suggestions to customers. Early in 2013, more than 85 \% of Amazon sales revenue was related to the personalized recommendation technology\footnote{https://www.mckinsey.com/industries/retail/our-insights/how-retailers-can-keep-up-with-consumers\#}.

Empowering product detail pages with smart recommendation engines is becoming common in fashion and apparel e-commerce since they help open new revenue channels and boost up-selling and cross-selling opportunities. This trend has exploded particularly in the last few years with the boost of AI-driven recommendation technologies. Nowadays, most leading fashion retailers have equipped their online platforms with advanced recommendation functions. ZARA, H\&M, MACY'S and ASOS\footnote{https://www.zara.com/, https://www.hm.com/, https://www.macys.com/, https://www.asos.com/} can now provide relevant recommendations on the product detail pages, which recommends similar and matching products, or even outfit suggestions. There have been many technological startups running fashion recommendation-relevant businesses. For example, Style Advisor from WideEye\footnote{https://wideeyes.ai/style-advisor/} is a solution that enables retailers to automatize the complete-the-look process and make this expandable to their entire catalog. It inspires shoppers with high-quality outfit recommendations and increases their shopping basket. Stylitics\footnote{https://stylitics.com/products/e-commerce/} provides a service of automatically displaying outfit ideas on the product page during customers' browsing. There are others including Suggestyny\footnote{https://www.suggestyny.com/} which offers style-specific fashion item recommendations and Easysize\footnote{https://www.easysize.me/} which recommends size and fit for apparel and shoes. As we can see, fashion recommendation has now become a critical technology that is ubiquitous in the whole fashion industry, especially the retailing section.

The prosperous industrial application is owing to the steady research development. The number of research publications on fashion recommendation has been continuously increasing over the years, providing strong evidence that this research area is rapidly developing. In particular, with the accumulation of fashion-related data, many conventional tasks in fashion domain now can be analysed from a data science or computing perspective. Fashion recommendation methods are usually developed based on a wide range of fundamental research~\cite{dubreuil2020traditional,silva2019big,acharya2018big,xu2023multi}. 
For example, for personalized fashion recommendations, Collaborative Filtering (CF) usually serves as a basis for the development of advanced methods that can effectively explore users' fashion preferences~\cite{yu2018aesthetic,vbpr}. 
For the complementary recommendation task, metric learning ideas and methods are commonly adopted to learn the compatibility space(s), in which compatibility between items can be measured and further used as the criteria for complementary recommendation~\cite{veit2015learning,vasileva2018learning}. In recent years, deep learning (DL) has shown impressive performance in many research fields. The influence of DL has been stellar in the field of information retrieval and recommender systems~\cite{bin2021entity,zheng2022asynchronous,zhang2019deep}, including fashion recommendation. DL-based methods permeate the whole of the fashion recommendation area. For example, for outfit recommendation, graph neural networks (GNNs)~\cite{kipf2016gcnsemi} and recurrent neural networks (RNNs)~\cite{hochreiter1997long} are widely applied to model the outfits consisting of multiple fashion items~\cite{li2020graph}. Also, Transformers have been adopted to model the user interaction with items in chronological order~\cite{chen2019pog}. While some advanced technologies have achieved great success in other recommendation contexts, they are still under-explored in fashion recommendation. For example, the Knowledge Graph (KG) technology is powerful to incorporate high-order and multi-relational connectivities between users, items and the other affiliated attributes with a hybrid structure~\cite{ji2021survey, wang2019kgat}, and causal reasoning is another promising tool for the recommendation for tacking several issues such as bias, fairness and explainability~\cite{xu2021causal,li2021towards,xu2020learning}. However, these technologies have not been explored much in the fashion domain so far. All these deep learning advances are expected to perform more important roles in future fashion recommendation research, which will be discussed in the remaining part of this survey.

\noindent \textbf{How does this survey differ from others?}
Due to the great significance of fashion recommendation to the academia and industry, it has attracted much research attention in the information retrieval, multimedia analysis, and computer vision communities. In recent years, workshops and tutorials on fashion recommendations have been held at top-tier conferences. For example, WWW 2019 held a tutorial on \textit{Concept to Code: Deep Learning for Fashion Recommendation}. SysRec, the specific conference about recommender systems, has been holding the \textit{recsys$\mathcal{X}$fashion} workshop every year since 2019. These evidences suggest that fashion recommendation has been becoming an independent and active area with great research value.

Even though fashion recommendation has been attracting increasing research attention, surveys or reviews specifically focusing on this topic are limited. 
The are several review papers on recommender systems in the literature~\cite{zhang2019deep, zangerle2022evaluating, deldjoo2020recommender, alhijawi2022survey, wu2020graph}, but them pay no special attention on the fashion domain. 
From another perspective, surveys focusing on computational fashion review fashion recommendation as a subset.~\cite{cheng2020fashion, gu2020fashion,song2018multimedia,wazarkara2020bibliometric,chakraborty2020comprehensive,mohammadi2021smart,feng2020become,chakraborty2021fashion}. 
For example, Cheng \myet~\cite{cheng2020fashion} categorize fashion recommendations into three groups, i.e., fashion compatibility, outfit matching, and hairstyle suggestion. They review each group of studies covering approaches, datasets, and evaluation. Gu \myet~\cite{gu2020fashion} review fashion recommendation and fashion compatibility as two independent research areas  belonging to high-level fashion applications. Despite these efforts in summarizing fashion recommendation research, they cover only a limited part of this research area and lack comprehensiveness and in-depth analysis. It is reasonable since fashion recommendation is not the main focus of these surveys. Currently, most fashion-related surveys from a computing perspective pay more attention to Computer Vision (CV) problems in fashion, while less to recommendation problems. There is also a survey that specifically focuses on apparel recommender systems~\cite{guan2016apparel}. However, it is from the perspective of design rather than technologies and algorithms. Another relevant survey published recently investigates clothing matching~\cite{zhang2020deep} and pays special focus on deep learning-based approaches. However, it is rather brief and covers only around 40 papers in total. Moreover, it only investigates fashion-matching recommendations while ignoring other sub-tasks.

From the above analysis, we can see that this paper differs greatly from all the previous surveys. To the best of our knowledge, this is the first survey that specifically focuses on fashion recommendations with advanced machine learning, data mining, and computer vision technologies. It aims to sort out the research area of computational fashion recommendation by organizing, reviewing, and summarizing relevant research papers and resources. Specifically, it groups existing studies into several sub-tasks and systematically reviews each sub-task, which has not been done in any existing work. Furthermore, this survey explores fashion recommendations in a more detailed, in-depth, and comprehensive manner by elaborating specific problems, solutions and limitations. It also discusses specific challenges of fashion recommendation and poses promising research directions for follow-on studies.

\noindent \textbf{How do we collect the paper?} We use the Google Scholar engine to search for relevant academic papers with several keywords including fashion recommendation, outfit recommendation, fashion compatibility, etc. We also explore high-profile conferences and journals in recent years to find relevant papers, including the ACMMM, KDD, WWW, CVPR, ICCV, ECCV, AAAI, SIGIR, ICDM, IJCAI, and TMM, to name just a few. Several papers published in workshops of top-tier conferences or pre-published in arXiv are also included. However, considering the reliability of the references and resources, this survey is mainly based on formal publications in well-recognized conferences and journals (most are listed above). We believe that there exist a lot more publications that are relevant to our topic in the literature, but summarizing and reviewing mainstream publications enable us to make sound analysis and discussion, which can better elaborate the development of the targeted fashion recommendation area. 

\noindent \textbf{Contribution of this survey.} The goal of this survey is to thoroughly review the literature on computational fashion recommendations. It provides a panorama through which readers can quickly understand the research area of fashion recommendation. 
The key contributions of this survey are as follows:

\begin{itemize}
    \item It reviews the fashion recommendation research under a proper categorization of tasks, each has a distinct problem formulation and research focus. Moreover, it summarizes experimental settings under this topic including model input, dataset, and evaluation protocols, which can help researchers new to this area to make an easy start to their studies. 
    \item It discusses the challenges and uniqueness of fashion, also how it differs from recommendations in general or other domains, which have been overlooked in most previous research papers.
    \item It systematically summarizes existing methods for each fashion recommendation sub-task, demonstrating its developing directions. On top of that, it discusses research limitations in each sub-task, offering a comprehensive survey on methodology.   
    \item  It emphasizes several important perspectives and remaining gaps in the research of fashion recommendation, and also points out promising directions for future work in this area. 
    \item It analyses the industrial needs and academic research outcomes, which facilitates discussions on how the fashion industry can really benefit from the development of recommendation and other relevant technologies. 
\end{itemize}

\noindent \textbf{Structure of this survey.} In the rest of this survey, we first give an overview of fashion recommendation in Section~\ref{sec:task}. In Section~\ref{sec:personalized_rec} to Section \ref{sec:special_rec}, we introduce and summarize the literature on specific research areas: personalized fashion recommendation, complementary recommendation, outfit recommendation, and other special fashion recommendations. We review each sub-task systematically by introducing the problem formulation and the state-of-the-art development. In Section~\ref{sec:dataset}, we introduce the important experimental settings in fashion recommendation, including the model input and evaluation protocols, as well as the existing datasets. Furthermore, we provide an in-depth discussion to provide more insights on this research topic and also propose potential future directions in Section~\ref{sec:future_direction}. Lastly, we summarize the whole survey and give the conclusion in Section~\ref{sec:conclusion}.
\section{Overview of Fashion Recommendation and Sub-tasks}
\label{sec:task}
Fashion recommendation generally refers to tasks that recommend fashion-related objects to certain or unspecified users. The research problem can be different when focusing on different aspects. As a result, studies on fashion recommendation are very diverse in the literature in terms of different research targets or problem formulations. As we know, recommendation is a type of high-level tasks in the field of fashion analysis and it is usually supported by low-level research such as clothing parsing, attribute recognition, and others~\cite{gu2020fashion,cheng2020fashion}. It is therefore formulated as a CV-related task at the early stage of its development~\cite{jagadeesh2014large, zhou2019fashion,simo2015neuroaesthetics} when most studies actually focus on extracting semantic information from fashion images. With fundamental CV problems gradually being addressed, it is recognized more widely that the user--item or item--item relationships, as well as their relationships with content information, are more important research focuses. To this end, fashion recommendation is more frequently formulated as an information retrieval (IR) or data mining (DM) problem. In this survey, our main attention is paid to research with IR/DM formulations. This section gives an overview of the fashion recommendation area by: 1) discussing its uniqueness; 2) introducing key terms and concepts in the context of fashion recommendation; and 3) demonstrating the categorization of specific tasks in this area.

\subsection{Uniqueness of Recommendation in Fashion Domain} In general, a recommender system~\cite{ricci2011introduction,jannach2010recommender} aims to provide personalized online product or service recommendations to users by predicting their interest in an item based on given information of the items and users, as well as the interactions histories. Applications of recommender systems include recommendation of movies, music, television programs, books, documents, websites, conferences, scenic tourism spots, learning materials, etc~\cite{diao2014jointly,van2013deep,peake2018explanation,kang2018recommendation,sohail2013book}. 
They cover the areas of e-commerce, e-learning, e-library, e-government, and e-business services~\cite{lu2015recommender}. 
From this perspective, fashion recommendation can partly be seen as a specific application of recommender systems in e-commerce. There is no doubt that fashion is personalized, as most recommendation systems are designed for~\cite{vaccaro2018designing}. But beyond that, fashion recommendation may also need to consider fashionability or compatibility in different situations. The purpose of recommendation in fashion is not only to help users with over-choice~\cite{zhang2019deep}, but also to provide styling or dressing advice for them~\cite{guan2018enhancing,guan2016apparel,dahunsi2021understanding,barnard2020fashion}. For example, for the complementary recommendation task, the main research problem is to model the compatibility between fashion items. There are more domain-specific challenges in fashion recommendation, which may result from the differences between fashion recommendation and recommendation in the general domain as summarized below:

\begin{itemize}
    \item Visual information matters more in the fashion domain than in most other domains. The very intuitive phenomenon is that one will not buy clothes or other fashion items without seeing them, physically or in a picture. Visual appearance plays an important role in the decision-making process of users when shopping for fashion items. Therefore, how to take good advantage of fashion product images to effectively leverage visual information is a key factor in fashion recommendation~\cite{yu2018aesthetic}.
    
    \item Attributes are rich and fine-grained. Usually, fashion items have abundant attributes, many of which can influence the users' appreciation of the items greatly. Important attributes such as styles and colors have been explored in previous studies and found to be effective in helping improve the recommendation performance.~\cite{nguyen2014learning}. However, on the other hand, fashion attributes are detailed and not easy to effectively incorporate, which is another main challenge in fashion recommendation. 

    \item Personalization is not the only criterion. Fashion recommendation studies not only the personalized fashion preference of users, but also the compatibility level between fashion items, or even the fashionability of single products. Since the research goal can be different from general recommendations, the problem formulation can be different and diverse when focusing on different aspects. 
    
    \item Exceptional data sparsity. Although sparsity is a widely acknowledged challenge in recommender systems, it is even more significant in the fashion domain as the item set is so huge. Svendsen has pointed out that the principle of fashion is NEW~\cite{svendsen2006fashion}. Unlike books or music, fashion items are updated very fast and constantly accumulates. Each item may have only been chosen by very few or even no users, which makes the interaction density of items very small. Therefore, existing methods, such as CF, that count on modeling the interactions between the user and item may not be less effective in the fashion domain~\cite{hu2015collaborative}.

\end{itemize}

\subsection{Terms or Concepts Frequently Used in Fashion Recommendation} Some terms are frequently used in this survey and other fashion recommendation papers, which are briefly introduced here to help the readers understand this research area more easily without any confusion. In the fashion recommendation context, we usually have the \textbf{user}, who is the person we make the recommendation for. In the retailing context, a user usually refers to a customer while in other contexts such as fashion styling the user can be a client. The \textbf{item} specifically denotes the fashion product, which could be an item of clothing, a pair of shoes, a handbag, or other fashion-related products. An \textbf{outfit} is usually composed of multiple items from different categories, such as the combination of a top, a bottom, and a pair of shoes. The \textbf{attributes} generally denote features of the user/item that characterize them from certain angles. For example, if the color of the item is white, the \textit{color} is one attribute of the item, and the corresponding \textbf{attribute value} is \textit{white} in this case. Commonly explored item attributes include \textit{category}, \textit{brand}, \textit{color}, \textit{pattern}, \textit{price}, \textit{material}, \textit{size}, \textit{style}, etc. In terms of the user, possible \textbf{attributes} include \textit{occupation}, \textit{body shape}, \textit{color}, etc. Taking \textit{occupation} as an example, the corresponding \textbf{attribute value} can be a \textit{teacher}. Other user attributes include \textit{body shape}, \textit{age}, \textit{color}, etc. However, due to privacy considerations, user attributes have been barely explored in existing research. The attributes of either the user or item are also called the \textbf{side information} (or associated/affiliated information) in many other recommendation papers~\cite{wang2017item}. \textbf{User preference/taste} generally refers to what styles of product he/she likes, which is usually subjective and abstract. In the recommendation studies, it is actually reflected by the user's appreciation of different items or outfits, and whether he/she selects the item/outfit or not, which is hard to explain explicitly. \textbf{Compatibility} is a term frequently used to describe the relationship between two fashion items and specifically measures whether the two items complement/match each other.

\begin{figure}
    \centering
    \includegraphics[width=\textwidth]{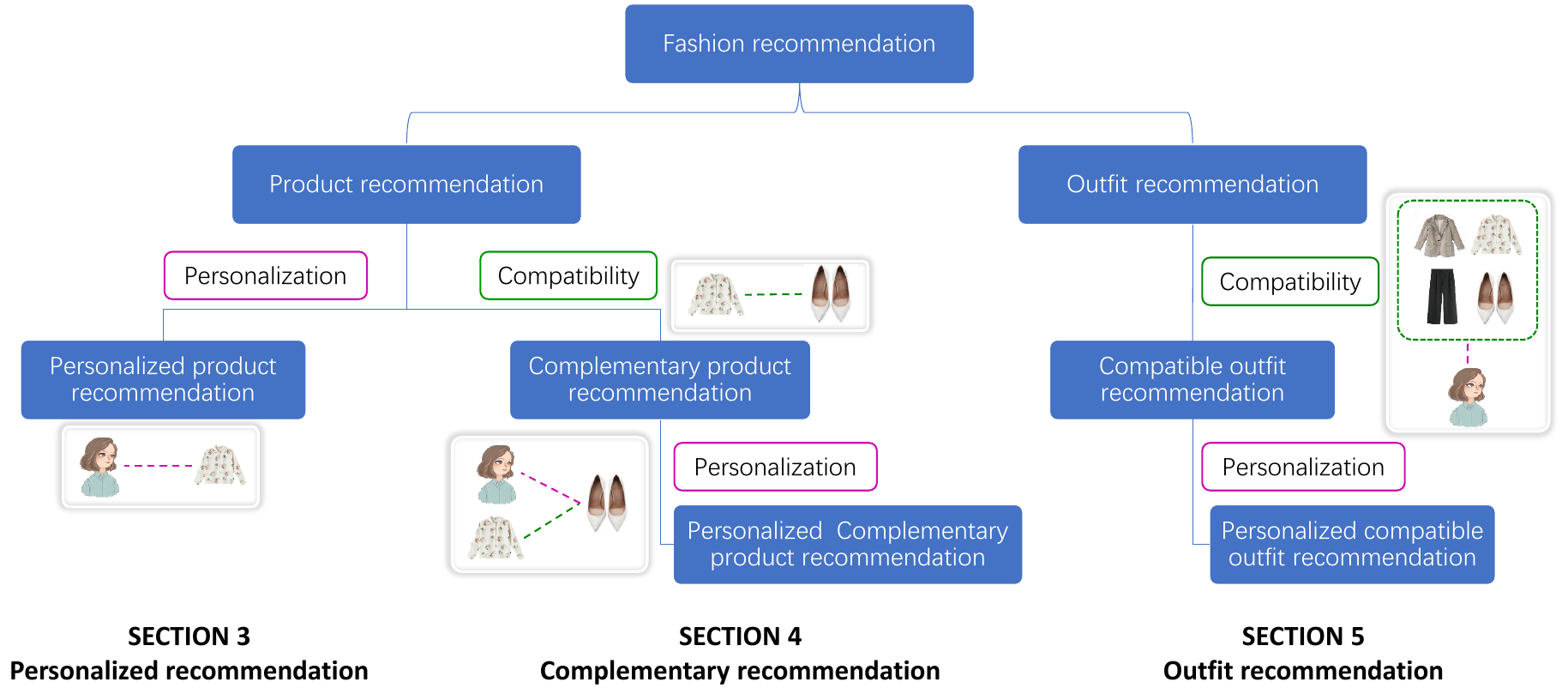}
    \caption{Categorization of fashion recommendation tasks. }
    \vspace{-10pt}
    \label{fig:task_cate}
\end{figure}

\subsection{Categorization of Fashion Recommendation Tasks}
In the literature, we can find a large number of papers working on the same topic of \textit{Fashion Recommendation} but focusing on quite different problems. For example, all the three papers~\cite{kang2017visually, hu2015collaborative, zhou2019fashion} have \textit{Fashion Recommendation} in their titles, but their problem formulation, proposed methods, experimental settings, including testing data and evaluation protocols are completely different. Specifically, Kang \myet~\cite{kang2017visually} work on the task to recommend fashion items for given users by exploring the personal preference of users. Hu \myet~\cite{hu2015collaborative} study how to effectively recommend fashion items for specific users, meanwhile to match with given items and make a good outfit. Zhou \myet~\cite{zhou2019fashion}, instead, study the task of recommending similar or compatible items for a given item. As we can see, the term \textit{Fashion Recommendation} refers to different research problems in different contexts or different research settings. Understanding the classification, or different variants, of fashion recommendation tasks is crucial for beginners in this area, which is also the preliminary objective of this survey.

The variety of fashion recommendation actually results from the broad concept of ``recommendation'', which can be interpreted in different ways. For example, tasks of recommending fashion items for certain users or to match with certain items can both be called fashion recommendation. As illustrated in Fig.~\ref{fig:task_cate}, based on what to recommend, fashion recommendation can be categorized into two groups, \textit{i.e.}, product recommendation and outfit recommendation. Fashion product recommendation generally refers to the tasks in which a single piece of fashion item is recommended. In comparison, fashion outfit recommendation aims to recommend a set of well-matched fashion items, which is more unique in the fashion domain. Considering the different recommendation requirements, product-level fashion recommendation can be further divided into personalized and complementary recommendations. The former focuses on personal preference modeling while the latter focuses on item compatibility modeling. Personalization modeling can also be involved in the complementary recommendation tasks to simultaneously explore the personal preference and item compatibility for more targeted functions. Different from product-level recommendation, fashion outfit recommendation is a more complicated and domain-specific task, which has attracted considerable research attention and produced a decent amount of literature. Outfit recommendation can also be divided into non-personalized and personalized varieties. The non-personalized outfit recommendation mainly works on the compatibility modeling of outfits, which focuses more at the level of set instead of pair. Similarly, the personalized outfit recommendation tries to explore additional outfit-level personal preference, which is usually beyond the simple combination of product-level preferences. 

\begin{table}[t]
    \centering
    \caption{Representative papers for various sub-tasks}
    \begin{tabular}{p{32mm}|p{43mm}|p{40mm}|p{27mm}}
    \hline
         \makecell[c]{Personalized Fashion\\ Recommendation \\(PFR in Section~\ref{sec:personalized_rec})}
         &\makecell[c]{Complementary Fashion\\Recommendation \\(CFR in Section~\ref{sec:compatibility_model})}
         &\makecell[c]{Fashion Outfit\\ Recommendation \\ (FOR in Section~\ref{sec:outfit_rec})}
         &\makecell[c]{Special Fashion\\ Recommendation \\(SPR in Section~\ref{sec:special_rec})} \\
         \hline
         \cite{vbpr,hou2019explainable,he2016ups,kang2017visually,liu2017deepstyle,mcauley2015image,nguyen2014learning,yu2018aesthetic,chong2020hierarchical,ding2021modeling,ding2021leveraging,he2016sherlock,chen2019personalized,bracher2016fashion,ding2022modeling} 
        & \cite{iwata2011fashion,shih2018compatibility,dong2020fashion,li2020fashion,liu2018learning,cucurull2019context,he2016learning,yang2021attribute,liu2019end,lin2019improving,lin2018explainable,laenen2020comparative,zhou2018fashion,tangseng2020toward,chen2018dress,yang2019interpretable,yang2019transnfcm,zhao2017deep, zhou2019fashion,song2018neural,song2017neurostylist,sun2020learning,vasileva2018learning,veit2015learning,yang2020learning,huynh2018craft,liu2020mgcm,song2019gp,polania2019learning,jagadeesh2014large,sun2020learning,tan2019learning,Liao2023recommendation} 
        &\cite{chen2019pog,cui2019dressing,dong2019personalized,feng2019interpretable,hsiao2018creating,hu2015collaborative,jiang2018outfit,lin2020outfitnet,liu2020learning,nakamura2018outfit,tangseng2017recommending,li2019coherent,li2017mining,simo2015neuroaesthetics,verma2020fashionist,wang2019outfit,zheng2020personalized,han2017learning,feng2018interpretable,jing2019low,singhal2020towards,li2020hierarchical,li2020fashion,lu2019learning,Sarkar2022outfittransformer,yang2020learning2,zhan2021a3,li2020learning}
        & \cite{dogani2019learning,hidayati2018dress,hsiao2020vibe,hsiao2019fashion++,karessli2019sizenet,liu2017weather,misra2018decomposing,otieno2007fit,peng2014personalised,sattar2019fashion,sembium2017recommending,sheikh2019deep,shen2007gonna,zhang2017trip,liu2018learning,lu2019learning} \\
         \hline
    \end{tabular}
    \label{tab:paper_classification}
\end{table}

There are other existing fashion recommendation studies tackling additional challenges on top of the typical user preference or item compatibility modeling. Some of them may focus on the recommendation task with special requirements or more detailed factors, such as the occasion-based / weather-based fashion recommendation or size recommendation. 
In Table~\ref{tab:paper_classification}, we summarize the representative papers addressing different sub-tasks, including \textbf{P}ersonalized \textbf{F}ashion product \textbf{R}ecommendation (\textbf{PFR}), \textbf{C}omplementary \textbf{F}ashion \textbf{R}ecommendation (\textbf{CFR}), \textbf{F}ashion \textbf{O}utfit \textbf{R}ecommendation (\textbf{FOR}) and \textbf{S}pecial \textbf{F}ashion \textbf{R}ecommendation (\textbf{SPR}).

\vspace{10pt}
\section{Personalized Fashion Recommendation}
\label{sec:personalized_rec}

\textbf{P}ersonalized \textbf{F}ashion product \textbf{R}ecommendation (\textbf{PFR}) aims to provide personal fashion suggestions for users, in most cases, suggestions about fashion products~\cite{kang2017visually,he2016ups,he2016sherlock}. The goal of PFR is similar to that of personalized recommendation in the general domain, which is to explore personal preferences of users from their historical behaviors. Therefore, solutions for general personalized recommendation can be applied to address similar problems in the fashion domain. Nevertheless, PFR has its uniqueness in several aspects: 1) the visual features are very important to describe fashion items; 2) items that a user interacts with, together or at different times, are likely to be related; 3) fashion products are usually seasonal and trend-aware, so customers' general preferences change obviously over the seasons; and 4) the total number of fashion items is usually very large and long-tailed in most cases. New items are constantly emerging, which makes the data sparsity issue even more severe. Because of its uniqueness along with challenges, PFR has attracted special research attention in the literature. 

\subsection{Problem Formulation} 
Letting $\mathcal{U}$ and $\mathcal{I}$ denote the sets of user and item, containing $N^u$ users and $N^i$ items respectively. Each user $u$ is associated with an item set $\mathcal{I}^{+}_u$ with which $u$ has interacted. In addition to those implicit feedback data, there is usually content information for the description of items, such as the images $\mathcal{X} \in \mathbb{R}^{N^i \times L \times W \times H}$. The goal is to train a recommendation model $f$ which can predict the personalization score of any item $i$ for a given user $u$, \textit{i.e.}, $s_{u,i} = f(u,i)$ based on the modeled user preference. Given a user $u$, by ranking the personalization scores of all candidate items $s_{u, i_1} > s_{u, i_2}, ..., > s_{u, i_{N^t}}$, the model can generate a recommendation list $[i_1, i_2, i_3, ..., i_{N^t}]$ specifically for $u$, in which the items that $u$ likes more rank higher. $N^t$ is the length of the recommendation list. 
\begin{figure}
    \centering
    \includegraphics[width=\textwidth]{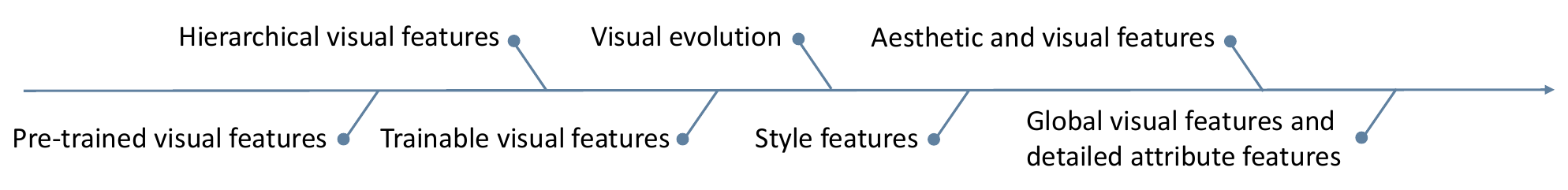}
    \caption{Major development of personalized fashion recommendation based on basic BPR framework by emphasizing visual information. Listed methods in the figure from left to right: \cite{vbpr}, \cite{he2016sherlock}, \cite{kang2017visually}, \cite{he2016ups}, \cite{liu2017deepstyle}, \cite{yu2018aesthetic}, \cite{hou2019explainable}.}
    \label{fig:person_develop}
    \vspace{-10pt}
\end{figure}

\subsection{General Methods}
Collaborative filtering is still the mainstream framework for PFR. Bayesian Personalized Ranking (BPR)~\cite{rendle2012bpr} is an algorithm that aims to optimize the personalized ranking based on the implicit feedback (viewing, purchasing, etc.) of users with the system. Combined with the most basic matrix factorization (MF) model~\cite{koren2015advances}, MF-BPR has been a benchmark personalized recommendation algorithm, being simple, easily applicable but effective in many recommendation scenarios. Specifically, MF models the probability of user-item interaction with the dot-product operation based on the user and item representation as follows~\cite{koren2015advances}:
\begin{equation}
    s_{u,i} = \alpha + \beta_u + \beta_i + \boldsymbol{e}^{T}_u \boldsymbol{e}_i,
\end{equation}
where $\alpha$ is the global bias, $\beta_u$ and $\beta_i$ are the bias specifically for $u$ and $i$. $\boldsymbol{e}_u$ and $\boldsymbol{e}_i$ are embeddings of the user $u$ and item $i$ respectively. The model is trained by the BPR algorithm in a pairwise ranking manner as:
\begin{equation}
    \text{BPR-opt} := \sum_{D}{\text{ln}\sigma(s_{u,i}-s_{u,j})-\lambda_{\Theta}||\boldsymbol{\Theta}||^2},
\end{equation}
where $D=\{u,i,j|i\in \mathcal{I}_u^+, j\in \mathcal{I}_u^-\}$, $\boldsymbol{\Theta}$ denotes all trainable parameters and $\lambda$ is the hyper-parameter to balance the weight of regularization. The rationale of MF-BPR is to rank those items catering to the user (having interacted) as high and those the user is not interested in (having not interacted) as low. In the definition of the training set $D$, $\mathcal{I}^+_u$ and $\mathcal{I}^-_u$ denote the positive and negative items set for $u$. In practice, $\mathcal{I}^+_u$ is usually the collection of items that the $u$ has interacted with, showing the user's interest in these items, which therefore is definite. However, the negative item set is not explicitly defined since users do not label items they are not interested in. In this case, the common assumption in here is that items not being interacted by the user construct a negative set for the user, i.e., $\mathcal{I}^-_u = \mathcal{I} - \mathcal{I}^+_u$.

Most existing PFR methods are developed based on the CF framework. However, since fashion is a domain in which visual factors are largely at play, most existing methods put effort into improving the incorporation of visual information to enhance recommendation performance. It is the common sense that \textit{"One wouldn't buy a t-shirt from Amazon without seeing the item"}~\cite{vbpr}, demonstrating the influence of the visual appearance of items on users' purchase decisions in fashion. From another perspective, incorporating visual factors of fashion products can also alleviate the \textit{cold start} issue~\cite{vbpr,kang2017visually, he2016ups}, which is a serious problem in the fashion domain than in others as new fashion products are emerging more frequently as mentioned before. With these considerations, \textbf{the development of PFR methods mainly goes along with the development of visual representation incorporation (as shown in Fig.~\ref{fig:person_develop})}, which includes several main stages as follows:

\begin{itemize}
    \item \textbf{Bayesian personalized ranking with visual features}. He \myet~\cite{vbpr} firstly propose the visual personalized ranking model (\textbf{VBPR}) to uncover visual and latent factors simultaneously by introducing the visual factors of users and items to the user--item interaction prediction as follows:
    \begin{equation}
        s_{u,i} = \alpha + \beta_u + \beta_i + \boldsymbol{e}^{T}_u \boldsymbol{e}_i + \boldsymbol{\theta}^T_u \boldsymbol{\theta}_i.
        \label{eq:vbpr}
    \end{equation}
    
As shown in Fig.~\ref{fig:person_case} (a), they project the high-dimensional CNN feature of fashion items to lower-dimensional ``visual rating" space to obtain the visual factor of $i$: $\boldsymbol{\theta}_i = \mathbf{W}\mathbf{v}_i$, where $\mathbf{v}_i$ is the CNN features and $\mathbf{W}$ is the projection matrix. Experimental results show that the recommendation performance was boosted with the introduction of visual factors. The authors also suggest that such performance boost is more significant on the clothing dataset than other categories such as phones, further demonstrating the special importance of visual information to the fashion domain. In another work, He \myet~\cite{he2016sherlock} propose to enhance the visual representation by encoding CNN features with different category-aware projections: $\boldsymbol{\theta}^{(c)}_i=\mathbf{W}^{(c)}\mathbf{v}_i$. They use a hierarchical embedding architecture which accounts for both high-level (colorfulness, darkness) and subtle (casualness) visual characteristics simultaneously, therefore achieving a better performance.

\begin{figure}
    \centering
    \includegraphics[width=0.95\textwidth]{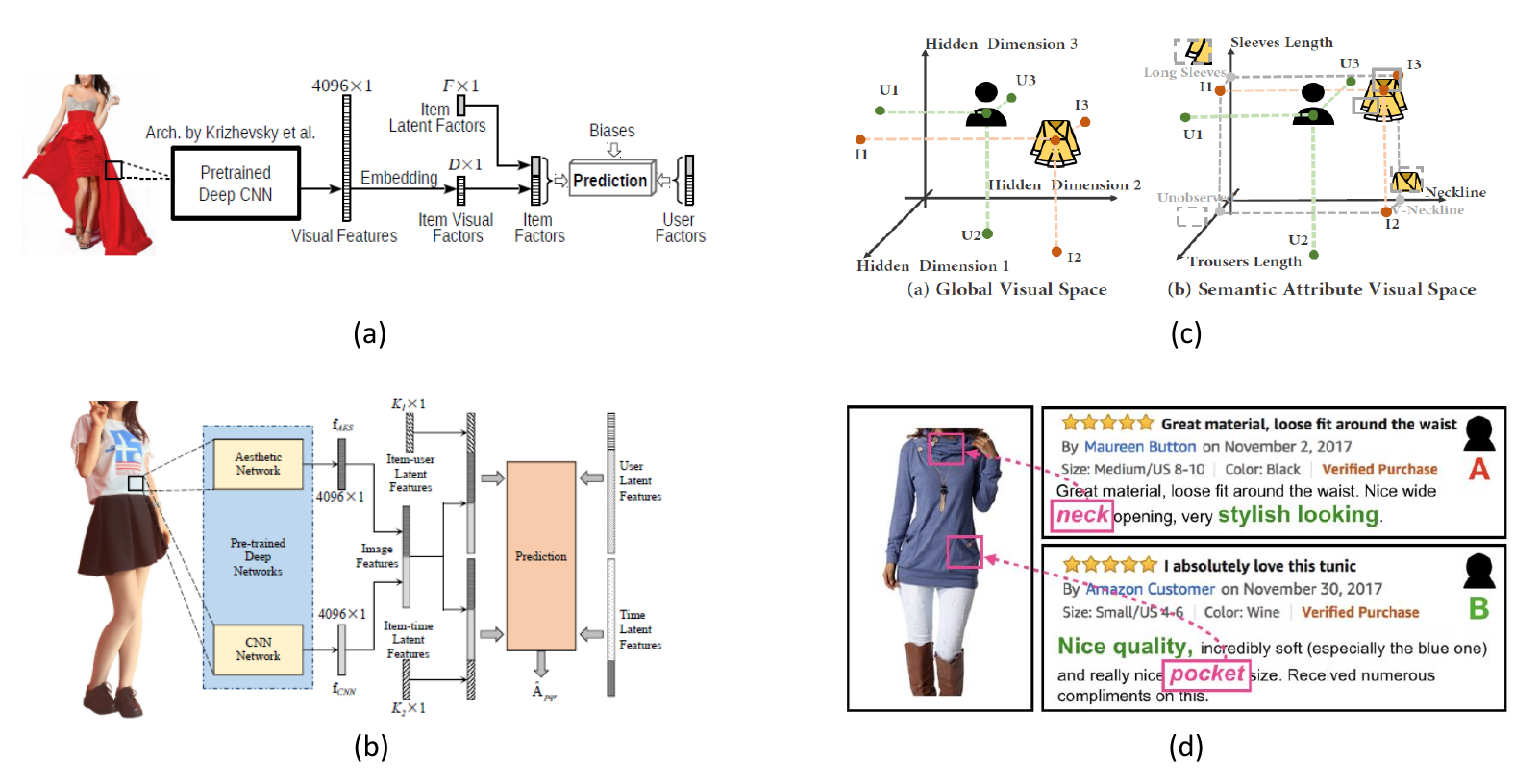}
    \caption{Representative personalized fashion recommendation models. (a) Visual Bayesian Personalized Ranking (VBPR)~\cite{vbpr}; (b) Introducing aesthetic features~\cite{yu2018aesthetic}; (c) Learning semantic attribute visual space~\cite{hou2019explainable}; (d) Bridging user comments with visual regions~\cite{chen2019personalized}.}
    \vspace{-15pt}
    \label{fig:person_case}
\end{figure}
    
\item \textbf{VBPR with trainable features}. To improve the incorporation of visual information, Kang \myet~\cite{kang2017visually} propose to train the image representation and recommender system jointly rather than directly applying the pre-trained visual features to the recommendation models. The experimental results in their work show that the introduction of the generative image model facilitates the recommendation performance on several clothing recommendation datasets significantly. From another perspective, He \myet~\cite{he2016ups} propose to consider the visual evolution in the recommendation model, trying to make the model not only visual-aware but also fashion-aware. Specifically, they build scalable models to capture temporal dynamics to help estimate users' personalized rankings with a one-class collaborative filtering framework.
    
\item \textbf{VBPR with fashion-related features}. The aforementioned methods encode fashion images into general visual representations, which is criticised for lacking fashion-specific information, such as aesthetics. Yu \myet~\cite{yu2018aesthetic} claim that when purchasing fashion products, consumers are not only concerned with ``what is the product", but also with ``is the product good-looking". Therefore, they propose to incorporate aesthetic features into the VBPR framework (Fig.~\ref{fig:person_case} (b)), which are extracted from an additional aesthetic visual assessment model. Experimental results demonstrate the performance improvement of the recommendation model, proving the importance of aesthetic modeling in fashion recommendation. Similarly, Liu \myet~\cite{liu2017deepstyle} propose to incorporate the \textit{style} features extracted by the DeepStyle model to enhance PFR. Instead of solely using global visual features, Hou \myet~\cite{hou2019explainable} propose to enhance the understanding of fashion items by incorporating attribute features. They introduce a semantic attribute space where each dimension corresponds to a semantic attribute and aim to learn users' fine-grained preference in various dimensions (Fig.~\ref{fig:person_case} (c)). According to the experimental results, this method achieves a significant performance improvement over the VBPR. It is also able to generate explanations for the recommendation results, which makes the recommendation more convincing. Likewise, in order to introduce more explanation to the recommendation results, Chen \myet~\cite{chen2018visually,chen2019personalized} propose to link user comments with specific visual regions in the image (Fig.~\ref{fig:person_case} (d)). They propose a model to explore the detailed interest of the users in terms of the visual part of the items to provide a reason for the recommendation. 
\item \textbf{Other efforts.} To address the cold-start issue, Bracher \myet~\cite{bracher2016fashion} propose to leverage both the visual features and ``expert labels" to describe items, modeling more semantic information, which is proven more effective for the cold-start recommendation settings. From the training perspective, instead of using the standard BPR framework, Chong \myet~\cite{chong2020hierarchical} propose a novel minimax ranking-based learning scheme for the visual-aware recommendation and achieved state-of-the-art performance. They take advantage of both the visual features and co-purchase data, simultaneously maximizing the preference discrepancy between positive and negative items and minimizing the gap between positive and co-purchased items.

\end{itemize}

\subsection{Sequential Methods}
Different from the methods in the previous subsection targeting personalized recommendation only, Ding \myet~\cite{ding2021leveraging, ding2021modeling} work on the sequential fashion recommendation problem, which considers the sequential order of users' interactions with items and therefore can make the recommendation happen in the right time. Compared to the non-sequential recommendation introduced above, it models the item--item transition in addition to the user--item interaction for exploring the patterns of user behaviors. To this end, Ding \myet~\cite{ding2021modeling} propose to specify the transition between items by introducing the instant user intents, which can be \textit{substitute}, \textit{match} or \textit{others}. In particular, they predicted the user's intent and based on the intent prediction results to predict the next item the user is likely to interact with. They leverage the intent modeling with basic translation-based sequential recommendation models~\cite{he2017translation}. In another work, Ding \myet~\cite{ding2021leveraging} try to exploit the high-level user--item and item--item connectivity based on the basic tensor decomposition-based model FPMC~\cite{rendle2010factorizing}, which effectively tackles the data sparsity problem and boost the overall recommendation performance.

\subsection{Summary}
As stated before, the visual factor is particularly important in the fashion domain. Therefore, existing methods of PFR emphasize the use of the visual information of fashion items and keep trying more effective ways to leverage it into the basic collaborative filtering frameworks. As illustrated in Fig.~\ref{fig:person_develop}, the development of this specific research field goes along with the development of visual representation incorporation, from the very preliminary general visual feature as fixed additional information~\cite{vbpr}, to the trainable global and detailed visual features~\cite{hou2019explainable}. The recommendation performance keeps increasing with this development. Research from other perspectives also makes contribution in other ways, such as enhancing the credibility of the recommendation with explanations~\cite{hou2019explainable} or improving the accuracy by taking more advantage of behavioral correlations~\cite{ding2021leveraging}.

Despite the achievement of the existing research, there are still limitations in this field. The first limitation is in only focusing on visual features while ignoring other factors that are also unique and important in the fashion domain, such as seasonality. On top of that, fashion trend is another important aspect that has been overlooked in previous research, yet has a great impact on the recommendation performance~\cite{ok2019recommendation,ding2021leveraging,ma2020knowledge,ding2022leveraging}. Moreover, exploring more dependencies and relations between user behaviors, even different types of behaviors (click or purchase), to better analyze and understand users would also be a promising direction for the next step in the development of PFR. 
\section{Complementary Fashion Recommendation}
\label{sec:compatibility_model}
\textbf{C}omplementary \textbf{F}ashion \textbf{R}ecommendation (\textbf{CFR}) refers to the task of recommending compatible fashion item(s) to match with a given one, which is one of the mainstream research tasks in the field of fashion recommendation. In some papers, this task is also named fashion coordination/coordinates recommendation or mix-and-match recommendation~\cite{iwata2011fashion, gu2017fashion,kang2019complete}. The criterion for CFR is the compatibility between items. In the context of fashion analysis, compatibility measures whether two items in different categories (rarely in the same category) match/complement each other. Since compatibility is an abstract concept and difficult to be defined, mainstream CFR methods try to explore it from a large amount of available item-matching data. Such a task is named \textbf{compatibility modeling}, which is exactly the key research problem for CFR tasks~\cite{vasileva2018learning}, also the main focus in this section of the review. 

In general, a complementary recommendation is made according to the compatibility measurements between items. However, in reality, the quality of item matches is subject to the appreciation of people who are choosing or wearing them. In other words, personalization could play a role in the CFR, posing the \textbf{P}ersonalized \textbf{C}omplementary \textbf{F}ashion \textbf{R}ecommendation (\textbf{PCFR}). It integrates the modeling of both item compatibility and personal preference, therefore is more complicated, and has been relatively less studied so far. In this section, we cover more on the non-personalized CFR since it is one of the basic problems in fashion tasks and lays the foundation for many other high-level fashion recommendation tasks. For PCFR, we make brief reviews and discussions.

\subsection{Fashion Compatibility Definition} 
\label{subsec:compatibility_definition}
As mentioned above, fashion compatibility represents how compatible when two or multiple items are together, in other words, whether the combination of items makes a good match/outfit. Since compatibility is a subjective concept, items matched well to one person might be not to another person. However, from the data perspective, if two items are frequently combined together in high-quality outfits, they can be deemed as compatible, otherwise not~\cite{han2017learning, vasileva2018learning, song2017neurostylist}. The way of items \textit{being combined together} can be different in different situations. For example, users select several items to form an outfit, which usually happens in fashion community platforms such as Polyvore, or users just wear multiple items together, which can be found on social media. On top of these common situations, we can find another definition of compatibility posed in the Amazon Styles and Substitutes dataset~\cite{mcauley2015image}. It defines two compatible items $i$ and $j$ as (1) $i$ and $j$ are frequently bought together or (2) customers who bought $i$ also bought $j$~\cite{veit2015learning}. This definition also follows the logic that two items \textit{being combined together} are compatible, except here \textit{being combined} is based on the users' purchase or click behaviors. However, in the fashion domain, compatible items should be those that are worn together well instead of bought together. Therefore, this definition is deemed to be less appropriate for the fashion domain, which has also attracted relatively less follow-up research attention. In this section, our review covers both definitions but focuses more on the first one because it is more reasonable and specific for the fashion domain. 

\subsection{Problem Formulation}
Let $\mathcal{I}$ be the item set with $N^i$ items. Each item $i \in \mathcal{I}$ has a set of compatible items $\mathcal{I}^+_i$, in which each item $j^+$ is well matched with $i$. The goal of compatibility modeling is to build a predictive model $f$ that estimates the compatibility score $s_{ij}=f(i, j)$ between any two given items ($i$ and $j$). For any item $i$, an effective model $f$ should be able to predict the score between $i$ and its compatible item $j^+ \in \mathcal{I}^+_i$ higher than that with its incompatible item $j^- \in \mathcal{I}^-_i$, and $\mathcal{I}^-_i = \mathcal{I} - \mathcal{I}^+_i$. A CFR model can therefore recommend proper items for a given item $i$ by ranking the compatibility scores of all candidate items. If personalization is considered, user information and user--item interactions are required. Let $\mathcal{U}$ be the user set, in which each user $u$ is interacted with multiple well-matched item pairs $(i, j)$, producing a set of triplets $\mathcal{T} = \{(u, i, j) | \mathcal{Y}_{(u,i,j)} = 1\}$, $\mathcal{Y} \in \mathbb{R}^{N^u \times N^i \times N^i}$ represents the user-item pair interaction tensor. The target of PCFR is to train a model $g$ able to score any triplet $s_{u,i,j} = g(u, i, j)$ based on the personalization and compatibility measure. Given a user $u$ and an item $i$, an effective model should score the suitable item $j^+$ higher than unsuitable ones $j^- $, \textit{i.e.}, $s_{u,i,j^+} > s_{u,i,j^-}$, where $(u,i,j^+) \in \mathcal{T}$ and $(u,i,j^-) \notin \mathcal{T}$.

\subsection{Complementary Fashion Recommendation Methods} The key to achieving high-quality complementary recommendation is the effective modeling of compatibility between items. In order to make a clear introduction, we categorize existing methods for compatibility modeling into three main groups: 1) topic model-based methods; 2) metric learning methods; and 3) graph-based methods\footnote{We focus on pairwise compatibility modeling in this section and leave the outfit-level compatibility to the next section.}. It needs to mention that such a method categorization covers the majority of existing methods in this sub-area but still leaves a few that do not belong to any one of the three, which we will also introduce as other methods at the end of this sub-section. Moreover, many methods are developed based on more than one basic technology or idea, our categorization is based on the main paradigm that each method follows.

\subsubsection{Topic Model-based Methods}
\begin{figure}
    \centering
    \includegraphics[width=\textwidth]{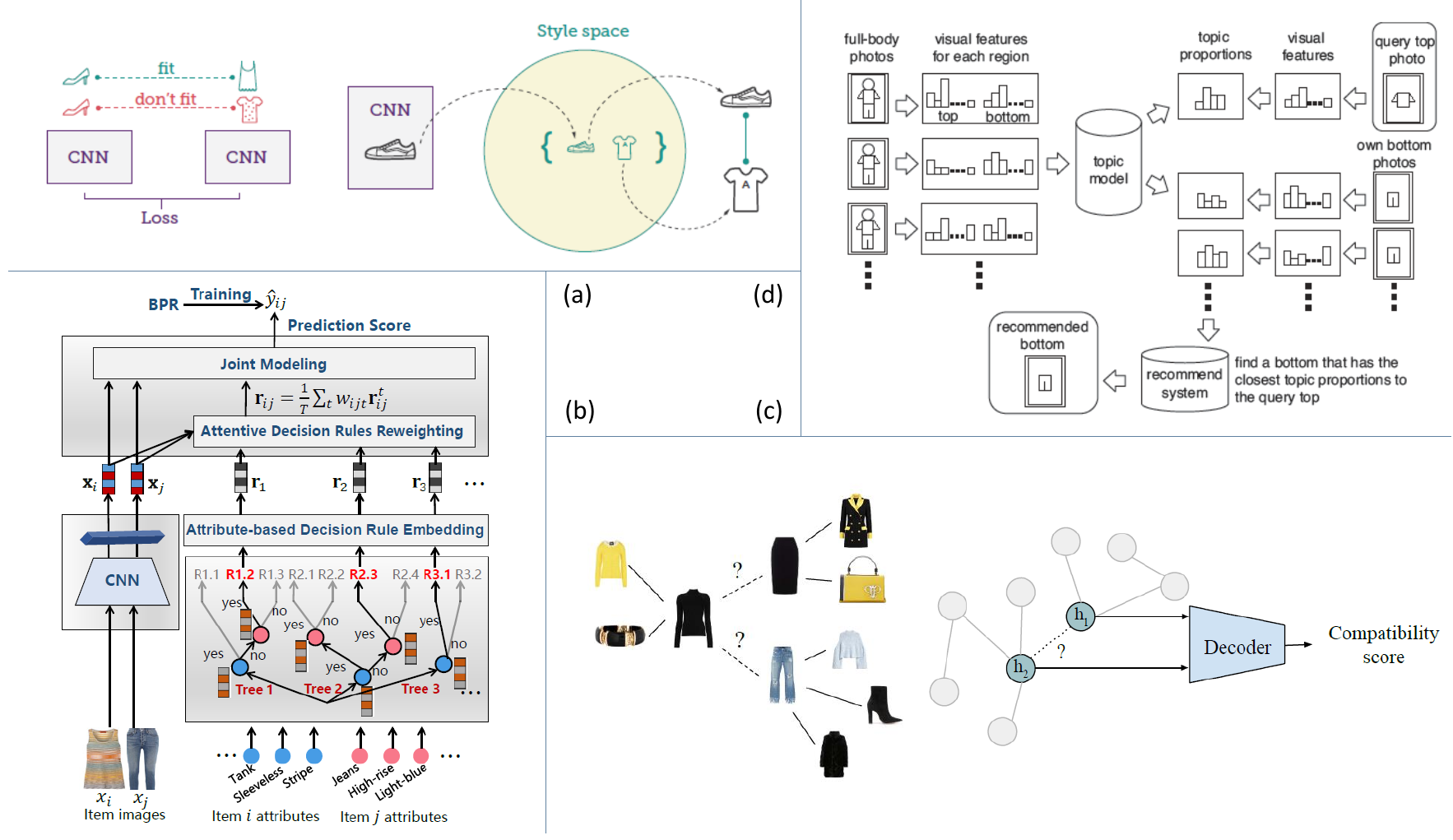}
    \caption{Four representative methods for pairwise fashion compatibility modeling. (a) Learning global compatibility space~\cite{veit2015learning}; (b) Applying decision tree~\cite{yang2019interpretable}; (c) Graph-based method to leverage context information~\cite{cucurull2019context}; (d) Topic modeling method~\cite{iwata2011fashion}}
    \vspace{-5pt}
    \label{fig:compatibility_methods}
\end{figure}

Early-stage studies address the compatibility modeling problem with topic models, specifically, the adapted Latent Dirichlet Allocation (LDA) methods~\cite{iwata2011fashion, jagadeesh2014large, zhou2018fashion}. LDA is a probabilistic generative model which has been successfully used for a wide variety of applications including information retrieval, collaborative filtering, and image modeling. It has also been applied in the fashion domain for other relevant tasks such as style analysis~\cite{vaccaro2016elements, hsiao2017learning}. From the perspective of topic modeling, a fashion outfit can be seen as a \textit{documents} covering several main \textit{topics} and attributes can be seen as \textit{words}. As shown in Fig.~\ref{fig:compatibility_methods} (d), the idea is that well-matched item pairs will have a certain topic proportion, meaning different items should have close topic proportions to make a good match together. Iwata \myet~\cite{iwata2011fashion} proposed to use LDA to learn co-occurrence information about visual features in top and bottom regions in full-body street photos to capture the coordinates. They tested the proposed method with three relevantly small datasets, which has been criticized as not being enough to ascertain the effectiveness of the method~\cite{jagadeesh2014large}. 

Combined with Gaussian Mixture Models (GMM), Jagadeesh \myet~\cite{jagadeesh2014large} learn topic patterns of different fashion items, aka, compatibility in a more efficient manner. When the model is well learned, given a query item, such as a shirt, the model firstly predicts the topic distribution and accordingly samples Markov Chain to generate new transformed samples to retrieve the nearest complementary items for matching. In comparison, Zhou \myet~\cite{zhou2018fashion} propose a more complicated method, which trains two hybrid topic models for the upper-body and lower-body clothes respectively to obtain the clothing representations, and then measures the compatibility with another collocation topic model. But the idea of a collocation topic model to capture compatibility patterns is quite similar to the previous two. We can also observe that with the development of relevant technologies, visual features used in these three representative methods keep upgrading, from color histograms to CNN features, which enhances the model from another perspective. However, these topic model-based methods apply the straightforward probabilistic model, which makes the performance rely more on the quality and choice of item features, \textit{i.e.}, the input of the model instead of the model design.

\subsubsection{Metric Learning Methods}
A large number of methods tackling the compatibility learning problem follow an established metric-learning paradigm~\cite{he2016learning,yang2006distance,suarez2021tutorial,kulis2012metric} with three main steps: (1) collect a large dataset of compatible (and non-compatible) items; (2) propose a parameterized similarity function; and (3) train the parameterized function to make compatible fashion items more `similar' than non-compatible ones. The `similarity' actually means compatibility, which is measured based on the learned representation of fashion items. Such representation space is therefore also called compatibility space. Following this paradigm, early-stage methods~\cite{veit2015learning} try to learn a general latent space based on visual features, in which the distance can measure the compatibility of any two fashion items, as shown in Fig.~\ref{fig:compatibility_methods} (a). However, such a simple one-compatibility space learning model may be less expressive, therefore, He~\myet~\cite{he2016learning} propose a multi-space learning model and adopt the mixtures-of-expert framework to obtain multiple notions for more comprehensive item features, which achieves more robust compatibility prediction. From another perspective, Lin \myet~\cite{lin2019improving} propose to learn the multi-modal compatibility between two items by combining the compatibility measure in each modal. Different from pairwise learning methods, Chen \myet~\cite{chen2018dress} extend traditional triplet neural networks to learn the compatibility patterns from well-collocated and bad-collocated samples, changing the common pairwise learning scheme into triplet-wise.

These early-stage methods treat \textit{compatibility} as a form of \textit{similarity}, which is not necessarily correct since \textit{compatibility} is \textit{NOT} transitive as \textit{similarity} is. For example, if item $i$ is close to both $j$ and $k$ when mapped into a general latent space, meaning $i$ is compatible with both items, $j$ and $k$ would be close too in the latent space. In that case, $j$ and $k$ would be compatible, which is apparently not true when $j$ and $k$ belong to the same category. To fix this limitation, type-specific space learning methods are proposed~\cite{vasileva2018learning,he2016learning}, which aims to learn spaces relating to item types/categories, or the type pairs. There are also cases where category information of items is not available. For these cases, Tan \myet~\cite{tan2019learning} propose a method of learning implicit substitute compatibility spaces to emphasize different semantic information (\textit{e.g.}, type pair). Such an idea is quite similar to He~\myet~\cite{he2016learning} mentioned above. They suggest that the learned implicit sub-spaces are able to capture compatibility between items specifically conditioned in certain semantic concepts, such as color or detailed design. The experimental results demonstrate a better performance compared to the basic Siamese Net-based method~\cite{veit2015learning}, as well as some of those learning explicit type-aware spaces~\cite{vasileva2018learning}.

\subsubsection{Graph-based Methods}
Graph techniques have been widely applied for fashion compatibility modeling for their power to model unstructured data and relationships in different ways. The first way is to use knowledge graph embedding learning techniques to model category-specific compatibility relations. Yang \myet~\cite{yang2019transnfcm} propose a translation-based method (TransNFCM) using TransE to model pairwise item compatibility (between item $i$ and $j$) based on the embeddings of items ($\mathbf{e}_i$ and $\mathbf{e}_j$) and a \textit{category-comp-category} relation embedding ($\mathbf{r}^{c_ic_j}$) as:
\begin{equation}
    d_{ij} = d(\mathbf{e}_i, \mathbf{r}^{c_ic_j}, \mathbf{e}_j) = ||\mathbf{e}^{c_i}_i + \mathbf{r}^{c_ic_j} - \mathbf{e}^{c_j}_j||^2_2.
\end{equation}
Following this work, Li \myet~\cite{li2020fashion} propose to apply TransR instead of the TransE operation and a new negative sampling strategy, which achieve better compatibility learning performance.

Another motivation for applying graph methods is to leverage more connectivity for the enhancement of the item representation learning~\cite{cucurull2019context, yang2020learning}. Most pairwise compatibility learning models only rely on item information and pairwise compatibility relation, while overlooking the context information, such as the connectivity with other items co-existing in the outfits. To effectively leverage the context information, Cucurull \myet~\cite{cucurull2019context} propose a model based on Graph Auto-encoder (GAE)~\cite{kipf2016gcnsemi} which applies convolutional message passing and information aggregation as shown in Fig.~\ref{fig:compatibility_methods} (c). Other than the matching (compatible) relationship, Yang~\myet~\cite{yang2020learning} propose to consider more heterogeneous relationships to incorporate richer extra-connectivity. Specifically, they encode the items and the connectivity using a graph modeling different relationships between items, such as \textit{purchased together by user} or \textit{also viewed by user} in this work. Experimental results validate the effectiveness of such a technical improvement. The additional relationships leveraged help to characterize the structural properties of items in a real-world e-commerce environment, and can thereby enhance the compatibility relationship modeling.

\subsubsection{Other Methods}
On top of the three main groups of methods, Yang \myet~\cite{yang2019interpretable} propose a non-metric method based on the decision tree. Their method enables the incorporation of rich attribute information of fashion items and connects item-level compatibility with specific attributes, offering an explicit and interpretable solution for the target problem. Boosted tree models are applied and defined as an ensemble of $T$ decision trees. They obtain the embedding of the $t$-th decision rule by combining the decision state embedding of all attributes:
\begin{equation}
    \mathbf{r}^t_{ij}: \vec{\mathbf{a}}^t_1 \rightarrow \vec{\mathbf{a}}^t_2 \rightarrow \cdot \cdot \cdot \rightarrow \vec{\mathbf{a}}^t_Z,
\end{equation}
where $\vec{\mathbf{a}}_z = \mathbf{a}_z + \mathbf{s}_z$ is the decision state of the $z$-th attribute that combines attribute embedding $\mathbf{a}_z$ and its decision embeddings $\mathbf{s}_z$, $Z$ is the total number of attributes. The proposed model manages to explicitly model the high-order attribute interactions. The final prediction is made by combining the visual and rule features (referring to Fig~\ref{fig:compatibility_methods} (b)). The experimental results show that this method achieves good accuracy and interpretability. Also targeting better interpretability, as well as improving the recommendation accuracy, Song \myet~\cite{song2018neural} propose to distill the matching knowledge from the matching dataset and inject the knowledge into the model to refine the compatibility space learning and achieve improved performance.

Another promising direction to improve complementary recommendation performance is jointly learning complementary recommendation models with other relevant tasks. For example, Dong \myet~\cite{dong2020fashion} propose to join the try-on task with the compatibility modeling, seeking to use the try-on appearance to guide and improve the outfit compatibility evaluation. Liu \myet~\cite{liu2020mgcm} employ generative adversarial networks (GANs) to sketch a compatible template as the auxiliary link between fashion items to enhance the item compatibility modeling with the item-template modeling. Moreover, Lin \myet propose to combine fashion image ~\cite{lin2019improving} and comment generation~\cite{lin2018explainable} respectively with the compatibility modeling task and prove these generation tasks can enhance the compatibility modeling through joint learning.

To better sort out the pairwise compatibility modeling research, we summarize representative papers in this area that have been published in recent years in mainstream conferences and journals in Table~\ref{tab:pairwise_comp} in Appendix, in which we categorize the method of each paper to different groups from the perspectives of input, dataset and evaluation setting. Please see Appendix for more information.

\subsection{Personalized Complementary Fashion Recommendation Methods}
PCFR is defined as the problem to recommend suitable fashion items for a given user \textbf{and} item. Since recommended items need to be compatible with the given item and meanwhile preferred by the given user, both compatibility and personalization play important roles in this task. Different from the methods reviewed in Section~\ref{sec:personalized_rec} and the above sub-section which only consider one-side requirement (personalized or compatible), PCFR needs to explore two types of relations from the data: user-item interaction and item-item compatibility. 
To address the problem, Song \myet~\cite{song2019gp} propose a VBPR-based framework that includes pairwise compatibility modeling between fashion items, with the personalized compatibility score as
\begin{equation}
     \vspace{-3pt}  
    s^{ui,j} = \mu s_{ij} + (1-\mu) s_{uj},
    \label{eq:gp-1}
    \vspace{-2pt}
\end{equation}
where $s_{ij}$ and $s_{uj}$ are compatibility and personalized score respectively. The personalized score $s_{uj}$ can be obtained in the same way as in Eqn.~\ref{eq:vbpr}. For the compatibility score, they use a similar dot-product operation to calculate based on multi-modal content features as:
\begin{equation}
    \vspace{-3pt}  
    s_{ij} = \pi(\mathbf{v}_i)^T\mathbf{v}_j + (1-\pi)(\mathbf{w}_i)^T\mathbf{w}_j,
\end{equation}
where $\mathbf{v}$ and $\mathbf{w}$ are visual and textural features for corresponding items, $\pi$ is the hyper-parameter to balance two parts, so as the $\mu$ in Eqn.~\ref{eq:gp-1}. A following work~\cite{Sagar2020paibpr} brought more interpretability to the GP-BPR model by leveraging more item attributes and attribute-level user-item and item-item relationship modeling. 

Despite the effectiveness, these methods only consider the linear combination of two objectives of modeling, \textit{i.e.}, personalization and compatibility, which is straightforward. In fact, from the experimental results of GP-BPR~\cite{song2019gp}, we can observe some imbalance between the two parts of modeling. Such a result seems reasonable since the two objectives to model are separate from different perspectives yet are coupling when observed in specific behaviors. Just like other tasks making predictions with multiple factors, PCFR needs to capture the personalization and compatibility patterns coupling in real-world data, which may raise many challenges, including but not limited to the imbalance learning problem.

\subsection{Summary}
CFR has experienced great development in the past few years and turned into one of the basic tasks in the computational fashion area. The key problem in CFR is modeling the compatibility between fashion items, which has been formulated in different ways and addressed with methods based on different types of techniques, such as metric learning, topic models, graph models, decision trees, and so on. After years of development along with advanced computational technologies, the performance of CFR methods have been generally improved.

However, most of the existing methods of compatibility modeling are completely data-driven and try to fit the fashion-matching data to explore the implicit compatibility patterns. Even though data-driven solutions provide new perspectives in fashion compatibility evaluation besides the traditional human-centered ways of evaluation, they are heavily dependent on the data, which might result in biased or bad results. For example, currently, the most commonly applied datasets are Amazon and Polyvore. For the Amazon dataset, we have mentioned in Section~\ref{subsec:compatibility_definition} that the definition of compatibility is not appropriate for the fashion domain. Based on this dataset, models can only learn to predict items that would get clicked or bought together, not get matched together. The other group of datasets, \textit{i.e.}, the Polyvore datasets, contains outfits generated by different online users, which are therefore not necessarily highly compatible as assumed. Under this circumstance, purely data-driven models can hardly explore good compatible patterns and make good recommendations.

In fact, there might be some general rules that fashion experts or professionals follow when they evaluate compatibility on item collocations, for example, the match of colors, categories, or printings. Such matching rules have also been used in computational fashion recommendation studies. Zhou \myet~\cite{zhou2019fashion} apply them to define matching categories and attributes so that they can make CFR based on item images, which turns the key research problem into accurate detection of categories and attributes. Such a solution has its limitations. On one hand, defining matching rules requires strong expertise in fashion, which can only be accomplished by fashion professionals with subjectivity. On the other hand, even though the defined rules are solid, they can only measure the compatibility between items from a single or partial perspective, unable to provide comprehensive compatibility measurements. Nevertheless, it offers inspiration that data-driven methods might be able to take advantage of this domain knowledge by introducing some solid rules. An existing attempt ~\cite{song2018neural} employs the teacher-student network scheme with an attentive knowledge distillation component to distill and incorporate matching rules to facilitate the neural network-based compatibility learning model.

From another perspective, most existing methods follow an implicit modeling scheme, which limits the explanations and persuasiveness of the recommendation models. Incorporating external domain knowledge maybe a possible solution to address it here. Besides, considering content-level compatibility~\cite{Sagar2020paibpr}, learning specific reasons for the judgment of compatibility~\cite{zou2020regularizing} or other methods are all promising and achieved good preliminary results, but still leave large possible spaces to explore. Moreover, most existing methods simply adopt some general solutions or bring more fine-grained features to the overall recommendation models to add interpretability. However, few of them exposed matching laws from a data perspective, which will make a real contribution from the computer science perspective to addressing the CFR problem.

\section{Fashion Outfit Recommendation} 
\label{sec:outfit_rec}
\textbf{F}ashion \textbf{O}utfit \textbf{R}ecommendation (\textbf{FOR}) generally refers to recommending a set of fashion items to users that match well with each other, which relies on effective outfit compatibility modeling. Depending on whether the personal tastes of the various users are considered, FOR can be further divided into non-personalized and personalized FOR. On top of outfit compatibility, \textbf{P}ersonalized \textbf{FOR} (\textbf{PFOR}) models users' personal preference towards outfits, which has attracted increasing research focus in recent years since personalization plays a key role in most fashion-related tasks. Unlike the personalized fashion recommendation reviewed above, which only explores the users' preferences towards a single fashion item, PFOR focuses on modeling users' tastes toward a combination of fashion items. In early research~\cite{simo2015neuroaesthetics,ding2019fashion}, outfits are treated as the minimum unit of analysis, and specific items in the outfit are not considered. For example, images of people dressed in outfits are frequently studied to investigate the fashionability or suitability of the outfits. More recently, FOR research incorporates item-level details. The user preference towards outfits is related to the specific preference towards single items in the outfit but beyond the simple combination of item-level preference.

\subsection{Problem Formulation} 
Let $\mathcal{U}$, $\mathcal{O}$, $\mathcal{I}$ denote the user, outfit and item sets respectively. Each outfit $o \in \mathcal{O}$ is a set of items $\{i^o_1, i^o_2, ..., i^o_{|o|}\}$, where $i^o_* \in \mathcal{I}$. Each user $u$ is interacted with a set of outfit $\mathcal{O}^+_u = \{o^u_1, o^u_2, ..., o^u_{|\mathcal{O}^+_u|}\}$, where $o^u_* \in \mathcal{O}$. Outfit compatibility modeling aims to train a model $f$ to measure the compatibility of a set of fashion items to tell whether they can make a good outfit. A common task setting is to predict the item for a subset of outfit $o$ with one item missing. Specifically, given $\hat{o} = \{i^o_1, i^o_2, ...i^o_{|o|-1}\}$, the model needs to predict the missing item by measuring and ranking the compatibility of the completed outfits filled with a given item $i$: $f(i|\hat{o})$. The task is to rank candidate item $i$ which is compatible with $\hat{o}$ higher than those that are not, which means $f(i|\hat{o}) > f(j|\hat{o})$, $i \cup \hat{o} \in \mathcal{O}$ and $j \cup \hat{o} \notin \mathcal{O}$. PFOR aims to recommend a list of outfits to a user, each consisting of a set of compatible fashion items. To this end, it needs to train a model $g$ that can predict the personalization score between the given user $u$ and the outfit $o$ as $s_{u,o} = g(u, o)$, measuring whether the outfit is preferred by the user. Since $o=\{i^o_1, i^o_2, ...i^o_{N_o}\}$, items in the outfit are treated as input of the function in many cases, turning the prediction model into $s_{u,o} = g(u, o, i^o_1, i^o_2, ...i^o_{n_o})$. An effective model should be able to predict a list of outfits $[o_1, o_2, ..., o_{N^t}]$ for a given user $u$, in which the outfits are ranked by personalization scores, consistent with the user preference.

\subsection{Outfit Compatibility Modeling Methods}
\begin{figure}
    \centering
    \includegraphics[width=\textwidth]{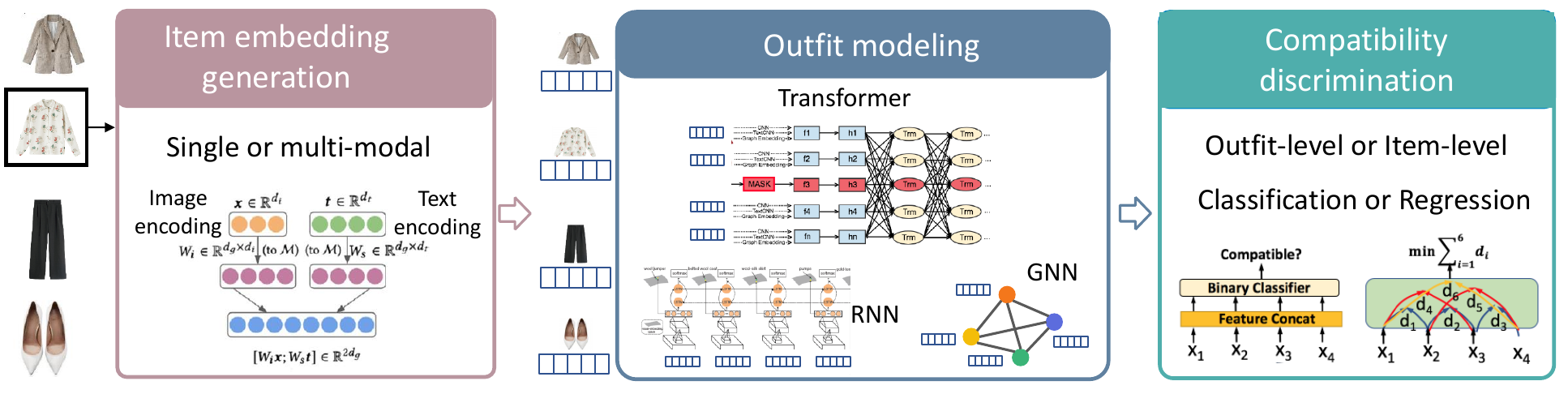}
    \caption{General pipeline for outfit compatibility learning, which roughly includes three steps: 1) item embedding generation, 2) outfit modeling based on item embeddings, and 3) outfit compatibility modeling.}
    \vspace{-10pt}
    \label{fig:outfit-compatibility}
\end{figure}

Outfit compatibility modeling can be seen as an extension task of pairwise compatibility modeling (the focus of Section~\ref{sec:compatibility_model}). It aims to predict the compatibility score for a given outfit, namely a set of items, which is also known as outfit composition evaluation. The general pipeline of outfit composition evaluation is to extract the effective feature for every item in the outfit and then calculate the overall compatibility score based on all item features. As shown in Fig.~\ref{fig:outfit-compatibility}, we detail the whole pipeline into three procedures: item embedding generation, outfit modeling, and compatibility discrimination. 

\subsubsection{Item Embedding Generation}
Item embedding generation, also called item feature extraction, aims to represent content information of items with effective representations, which can be achieved based on the either uni-modal or multi-modal item information. One common way is to extract multimedia features from visual and textual information. Many existing outfit compatibility modeling methods~\cite{wang2019outfit,cui2019dressing,han2017learning,lin2020outfitnet} adopt this method and apply Visual Semantic Embedding (VSE) technology~\cite{kiros2014unifying} to better integrate the two-modal features. Such an item embedding generation process is involved in the other two tasks too, \textit{i.e.}, personalized recommendation, and complementary recommendation, which therefore can be seen as a general problem in the area of fashion recommendation. We, therefore, discuss it in an independent part in Section~\ref{sec:dataset}. In this section, since \textbf{\textit{Outfit}} is the focus, we emphasise the outfit modeling methods in existing work.

\subsubsection{Outfit Modeling}
Outfit modeling specifically means the procedure that aggregates item-level representations to obtain the holistic representation of the outfit, which can be seen as a multi-instance pooling process. Different technologies can be applied to implement such a multi-object modeling and aggregation process. In existing work, the preliminary non-parametric feature concatenation, average pooling operation, and advanced neural network tools such as RNN, GNN, and Attention mechanism, have all been applied. Here we summarize existing work with regard to different outfit pooling methods they apply.   

\begin{itemize}
    \item \textbf{Non-parametric Pooling} methods are commonly used in different tasks for their merits of being feasible, easy-to-use, and efficient. In outfit modeling, non-parametric pooling methods such as \textit{concatenation}~\cite{tangseng2017recommending} and \textit{element wise reduction}~\cite{li2017mining} have been used. However, these simple aggregation methods may fail to model the complex dependency among outfit items, therefore would require more complicated techniques. 
    
    \item \textbf{Sequence Modeling} are widely applied for outfits. The rationale to apply sequence models is to learn the dependency between compatible fashion items when given a pseudo order among items. By modeling the whole ordered item sequence, the outfit representation can be obtained, which contains compatibility information of the outfit. If the sequence models are well-trained, they are able to predict missing items in the outfit or predict compatibility for any given outfit. Specifically, Han \myet~\cite{han2017learning} propose to apply the bidirectional Long Short-Term Memory (LSTM) model for the outfit, considering a collection of clothing items as a sequence with specific order. They input the features of each fashion item into the corresponding time step of the LSTM structure. This work contributes to the research field by effectively taking advantage of the representation power of LSTM, thereby providing a feasible solution to model the multiple items composition. Several follow-up papers propose variant approaches with similar RNN frameworks~\cite{nakamura2018outfit,jiang2018outfit,dong2019personalized}. For example, Li \myet~\cite{li2017mining} adapt RNN as a pooling model to encode variable numbers of fashion items and obtain the outfit representation. However, it has been acknowledged in recent studies that modeling fashion outfits as sequences are not appropriate as the items in the outfit should be order-less.
    
    \item\textbf{Graph Modeling} methods, instead, treat fashion outfits as unordered sets of items and model them with Graph based on the compatibility and other relations. As mentioned before, simply averaging pooling or other linear aggregation cannot exploit the complex relations among multiple items, while sequence modeling encodes some order information that does not exist. In comparison, graph modeling offers a more sophisticated way to explore dense and complex relations among items, therefore has attracted significant research attention in recent years and shown preferable performance. In specific, Cui \myet~\cite{cui2019dressing} construct a \textit{Fashion Graph} in which each node represents a category and each edge represents the interaction between two nodes. By putting each item into its corresponding node, an outfit can be represented as a subgraph of \textit{Fashion Graph}, and the task of modeling outfit compatibility can be accordingly converted to a graph inference problem. Specifically, they propose Node-wise graph neural networks (NGNN) to model the node interactions and learn node representation, further obtaining the outfit representation after item aggregation. Experimental results show that this performance is better than the LSTM-based model applied in Han \myet~\cite{han2017learning}. Similarly, Feng \myet~\cite{feng2018interpretable} construct an \textit{Outfit Composition Graph} taking item cluster center as nodes, which is achieved through the aggregation of color, shape, texture, and other item embeddings. The edges (connectivity between different clusters) are determined by the observed pairwise item-matching relations in different clusters. The final composition score of the graph can be inferred from the graph. Singhal \myet~\cite{singhal2020towards} build a type-conditioned graph autoencoder to learn a compatibility measure conditioned on type and context. They also introduce an attentive autoencoder that learns a style measure of compatibility through an outfit-level style representation. The whole model is trained with a reinforcement learning manner to learn a unified measure of compatibility. Liu \myet~\cite{liu2020learning} and Li \myet~\cite{li2020hierarchical} also apply the graph structure to model the outfits. Zheng \myet~\cite{Zheng2021collocation} proposed a disentangled graph learning method to explore implicit fine-grained compatibility between items. It also incorporates try-on analysis to evaluate the overall compatibility of outfits. To explore modality-specific compatibility, Song \myet~\cite{song2023modality} proposed the modality-oriented graph learning method. 
    
    \item \textbf{Attention Mechanism Modeling} is another feasible way to achieve unordered set modeling, applicable for the modeling of outfits and exploring the inherent outfit structures. Chen \myet~\cite{chen2019pog} adapt Transformer, a transduction model relying on self-attention to build the Fashion Outfit Model (FOM) for the modeling of multiple fashion items in the outfits. It learns outfit compatibility by the masked items prediction task, which inputs item set (outfit) with one position masked and predicts the item for it accordingly. 
    Sarkar \myet~\cite{Sarkar2022outfittransformer} embrace a similar idea of applying a Transformer encoder for outfit modeling and propose OutfitTransformer. Differently, for the outfit compatibility prediction task, they introduce an outfit token to capture a global outfit representation in the Transformer structure, which encodes the compatibility relationships among all the items in the outfit. Different from the above two methods borrowing the Transformer architecture, Yang \myet~\cite{yang2020learning2} propose a Mixed Category Attention Net (MCAN) that integrates category information to model the dependency and relations between items with the self-attention net (SAN). Li \myet~\cite{li2020learning} propose a Content Attentive Neural Network (CANN) that models the comprehensive compositional coherence of both the global contents and semantic contents among items in the outfit. The model is able to outperform RNN-based~\cite{han2017learning} and pairwise-based compatibility models~\cite{vasileva2018learning} for the outfit compatibility prediction task.
    
\end{itemize}

There are some other outfit modeling ways on top of the above four groups, for example, Multi-Layer Perceptron (MLP)~\cite{wang2019outfit}, but contribute quite a small portion. Since outfit modeling is quite a key technical part in the whole area of fashion recommendation, 
we summarize mainstream methods for the task and categorize them with different outfit modeling ways in Table~\ref{tab:outfit_comp} in Appendix. The corresponding input and evaluation settings for each work are also listed with the taxonomies given in Table~\ref{tab:input_eval_list} Section~\ref{sec:dataset}. From the summary we observe that early studies adopt non-parametric pooling or sequence modeling 
while the recent methods prefer graph and attention modeling, which have been discussed to be more appropriate and have better expressive ability. 

From the comparison between experimental results with different methods, the graph, and attention methods can generally achieve better performance compared to the former two in terms of outfit compatibility prediction or missing item prediction for outfits. Putting aside the improper outfit modeling with orders, the RNN models are powerful for modeling data with chronological dependency while the outfit structures are obviously not. Moreover, items can be related in various ways, but RNN models can only capture sharing outfit relationships, which may degrade its performance in this task. In comparison, graph or attention models are more flexible and powerful to model multiple types of item relationships as well as content information. However, one big concern is how to effectively train powerful models given limited data resources. From another perspective, these two types of methods need more computational space and time compared to the RNN-series methods and thus are less efficient. Moreover, when compatibility modeling is not the only objective in the task, such as PFOR, designing too complicated outfit modeling model might not worth it, sometimes just as effective as the average pooling operation based on our empirical observations.

\subsubsection{Compatibility Discrimination}
When the outfit structure is properly modeled, the next problem is how to predict its compatibility accordingly, which is usually dependent on the way to model outfits. For most sequence modeling/Transformer methods~\cite{han2017learning,jiang2018outfit,dong2019personalized,chen2019pog}, the compatibility of the whole outfit is mostly evaluated by the average of the probability of item in each position:
\begin{equation}
s_o = \frac{1}{n}\sum^n_{t=1}{\text{log}Pr(x_{t+1}|x_1,...,x_t; \Theta_f)},
\end{equation}
where $n = |o|$ is the length of outfit $o$, $Pr(\cdot|\cdot)$ is the conditional probability that has been produced by the sequence model, and $x_t$ denotes the output of the $t$ step. The specific calculation varies for different methods. For example, for Bi-LSTM models, two-directional predictions are included for the probability measurement~\cite{han2017learning}. In another work when category information is bonded with item representation for the outfit modeling, the joint distribution of the category and item tuple is considered to measure the outfit compatibility~\cite{yang2020learning2}. 
On top of sequential probability evaluation, outfit compatibility can be modeled based on the item-wise compatibility~\cite{cui2019dressing}:
\begin{equation}
    s_o = \sum^{|o|}_{i=1}{\sigma(f(\mathbf{e}_i)) \cdot \alpha(g(\mathbf{e}_i))}, 
\end{equation}
where $f$ and $g$ are two perception networks to model the weights and compatibility score for each item, $\sigma$ and $\alpha$ are leaky relu and sigmoid activation functions. The two parts in the multiplication model the importance of the item in the outfit and the compatibility level of the item respectively. Li \myet~\cite{li2017mining} treat the outfit evaluation as a binary classification problem and device a compatibility predictor on the outfit representation as:
\begin{equation}
   s_o = Pr(y_o=1|o) = \delta(W^Te_o+b),
\end{equation}
where $e_o \in \mathbb{R}^d$ is the representation of outfit $o$, $y_o$ is the ground truth outfit compatibility measurement of $o$, $y=1$ means compatible and $y=0$ means not.

\begin{figure}
    \centering
    \includegraphics[width=\textwidth]{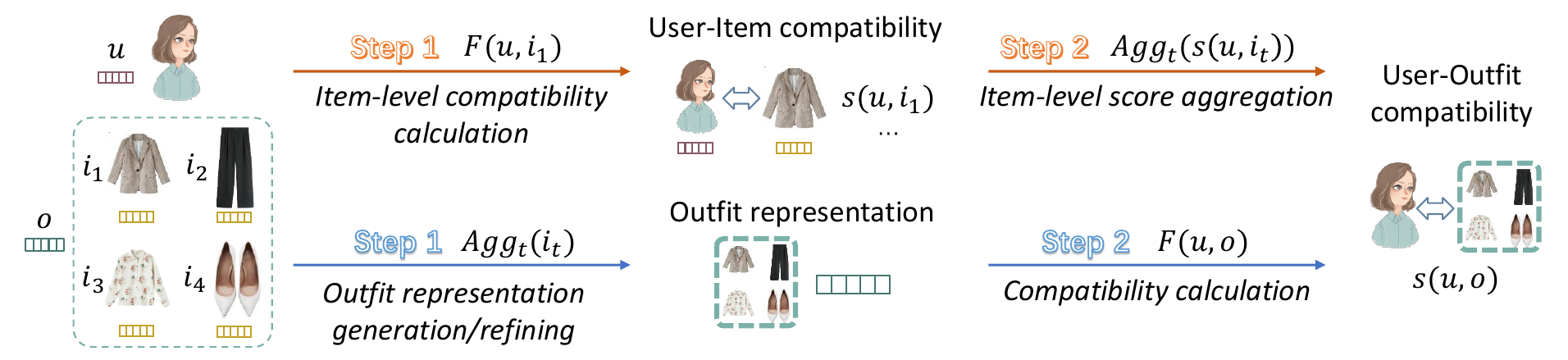}
    \caption{Two common pipelines (orange and blue) for personalized outfit preference modeling. The orange one models the user-outfit compatibility based on the item-level user preference while the blue one directly models that based on the outfit and user representations.}
    \vspace{-10pt}
    \label{fig:per-outfit-rec}
\end{figure}

\subsection{Personalized Outfit Recommendation Methods}
PFOR aims to recommend well-matched outfits to users that cater to their personal tastes. In the early stage, outfits are usually treated as the minimum unit to analyse. For example, Simo-Serra \myet~\cite{simo2015neuroaesthetics} make use of a Conditional Random Field (CRF) to evaluate the fashionability of fashion photos by analyzing the outfits (garment worn by people), people, and photo settings (background scene). They then make outfit recommendation based on the fashionability scores. Zheng \myet~\cite{zheng2020personalized} also investigates the personalized fashion recommendation problem solely based on the outfit images. With the area developed, current mainstream PFOR methods consider the intra-outfit item compatibility and the user's outfit preference. In other words, the compatibility of outfits and between users and outfits both need to be modeled. Existing approaches usually address the two modeling objectives separately. Outfit compatibility modeling has been introduced in the previous sub-section, leaving the rest of user-outfit compatibility modeling, which is equivalent to exploring the user preference for outfits. It is achieved in two general ways, as briefly illustrated in Fig.~\ref{fig:per-outfit-rec}. The first common way (the orange lines in Fig.~\ref{fig:per-outfit-rec}) evaluates the user preference for outfits based on the integration of item-level preference on the items~\cite{hu2015collaborative,dong2019personalized,lu2019learning,lin2020outfitnet} while the second way (the blue lines) obtains the outfit level representation based on item information first and then measures the user--outfit compatibility based on outfit and user representations~\cite{li2020fashion,zhan2021a3,lin2020outfitnet}.

Hu \myet~\cite{hu2015collaborative} conducted an early study on PFOR and propose a tensor factorization-based method, which basically follows the first pipeline shown in Fig.~\ref{fig:per-outfit-rec}.
The proposed method learns the personalized user preference for each item and the item--item compatibility in the outfit at the same time through functional tensor factorization. Lu \myet~\cite{lu2019learning} then apply a similar strategy to model the relationship between users, outfits, and items, except that they aim to obtain a binary representation of these objects to deliver better personalized outfit recommendation. For the personalized capsule wardrobe creation task, which can be formulated similarly to PFOR, Dong \myet~\cite{dong2019personalized} leverage the user body shape to the modeling of item compatibility among outfits, successfully train a model to provide body shape-aware PFOR. From another perspective, Lin \myet~\cite{lin2020outfitnet} propose an attentional fusion method that calculates the personalized item scores for the outfit first and then applies an attention pooling layer to generate the personalized outfit score. 

Following the second scheme of user-outfit modeling, Lin \myet~\cite{lin2020outfitnet} propose to fuse the item embeddings in the outfit with a personalized attention mechanism to generate the outfit embedding. They also conduct comparison studies for different settings with compatibility scoring at different stages, corresponding to the two pipelines we summarize in Fig.~\ref{fig:per-outfit-rec}. The experimental results show that generating outfit representations first and scoring based on user and outfit representations (the blue line pipeline in Fig.~\ref{fig:per-outfit-rec}) may achieve better personalized outfit recommendation performance than early scoring the user--item compatibility. However, the experimental results also suggest that evaluating the user--item compatibility first could be more helpful for the cold-start scenarios. Rather than generating outfit representation completely based on items, Li \myet~\cite{li2020hierarchical} assign each outfit an outfit id to take better advantage of the collaborative signals between users and outfits. To enhance the outfit representation, they apply message passing on the outfit--item graph to aggregate item information into outfit representation. They use the collaborative filtering strategy to model the user--outfit compatibility after obtaining an enhanced representation. On top of the common item information, a very recent work~\cite{zhan2021a3} leverages the knowledge graph with rich item- and outfit-level attributes to further enhance the outfit representation. They also exploit outfit information to enhance the user representation, which helps their method to achieve state-of-the-art PFOR performance.

\subsection{Outfit Generation Methods}
Fashion Outfit Generation (FOG) can be regarded as a pre-task of FOR, which is commonly defined in two forms: 1) generating the outfit images from scratch~\cite{rostamzadeh2018fashion,ak2019semantically,jain2019text,yu2019personalized}, and 2) making a collocation of items to generate the outfit. The first task focuses on image generation rather than outfit composition~\cite{liu2020learning,zhou2018fashion}. In this paper, we only discuss the second task because the first task is a complete CV problem, not the focus of this survey. 
The basic standard for FOG is completeness across categories and exclusiveness within the category, while the quality of the generated outfit is determined by the compatibility among the items~\cite{ding2023personalized}. 

It is formulated as a retrieval problem in many existing work. Chen \myet~\cite{chen2019pog} propose a transformer-based FOG model that encodes user preference in OG. They specifically generate the items in the outfit step-by-step in an auto-regressive manner, which consumes the previously generated items as input. Li \myet~\cite{li2019coherent} apply a similar iterative generation procedure. At each step, they rank all candidate items of the chosen type based on coherence and item--item compatibility. Lin \myet~\cite{lin2020outfitnet} first train a fashion compatibility model which can generate item embedding involving compatibility information. Then they interactively expand the outfit by adding items that are closest to the outfit.

A special form of FOG is the capsule wardrobe creation~\cite{hsiao2018creating, dong2019personalized}. Capsule wardrobe is a concept that uses a \textbf{minimal} quantity of clothing to create many outfits. Therefore, the task is to build a minimal clothing set from which items can be combined together and generate good-quality outfits. Hsiao \myet~\cite{hsiao2018creating} cast the capsule wardrobe creation as a subset selection problem which selects a subset from a large set of candidates that maximize compatibility and versatility. They define objective functions based on these two objectives and develop an EM-like iterative approach for optimization. In a further step, Dong \myet~\cite{dong2019personalized} propose to consider profiles in the model with a combinatorial optimization framework which jointly integrates garment and user modeling.

\subsection{Summary}
OR has experienced significant development in the last few years attributed to the research efforts from the information retrieval and multimedia communities. The task has been established and addressed in many different ways based on advanced neural network technologies. However, the research on FOR is still in its nascent stage and has several limitations.

First, the outfit compatibility modeling and PFOR tasks are tackled in isolation. In several studies, compatibility modeling serves as a pre-training process to generate an initial representation of fashion items for the main FOR task~\cite{chen2019pog}. Most state-of-the-art studies involve item-level compatibility modeling and outfit-level user preference modeling as two separate sub-tasks and evaluate their proposed models on these two sub-tasks independently.

Second, the evaluation for FOR is not quite solid so far. For example, Li \myet~\cite{li2020hierarchical} target the modeling of outfit-level user preference and apply the ranking evaluation on different methods, similar to general personalized fashion recommendation. However, such an evaluation does not suit generation-based FOR, in which the compatibility level of the generated outfit is not guaranteed. In the generation-based FOR by Chen \myet~\cite{chen2019pog}, online evaluation is applied, while it is not applicable to all studies, therefore obstructing the technical development to some extent.

Third, in existing FOG methods, the outfits are generated iteratively, for example, auto-regressively consuming a previously generated item as the input of the next~\cite{chen2019pog}. Such methods still assume that the outfit is in order, which is inappropriate, as pointed out in many papers~\cite{cui2019dressing, feng2019interpretable}. There is a need to develop better FOG models which can create the item set in parallel rather than one by one.

\section{Special fashion recommendation}
\label{sec:special_rec}
Faced with various application scenarios and requirements, several special research topics have also been posed and studied in the field of fashion recommendation, which will be introduced briefly in this section.

\subsection{Recommendation Based on Body Shapes}
Body shape has been widely ignored in fashion recommendation until recent years. It is an important factor that affects what garments will best suit a given person~\cite{hsiao2020vibe,hidayati2021dress,hidayati2021body}. In real-world applications, body shape-aware recommendation will be especially meaningful as people always want the clothes to fit their bodies well. There are two key problems in this task: 1) How to estimate the body shape of a given user and 2) how to make a proper recommendation for the estimated body shape. 

Hidayati \myet~\cite{hidayati2018dress} study the correlations underlying body shapes and clothing styles based on the celebrity body measurements and dressing images. They first build a body shape model which can classify the types of body shapes with the body measurements. After that, a body-style map is constructed to model the correlations between clothing styles and body shapes, to discover the dominant style for certain body shape types. In the following work, Sattar \myet~\cite{sattar2019fashion} expand the scope of recommendation from celebrities to ordinary people. They propose a multi-photo method to estimate the shape of each user and try to investigate the relationship between body shape and clothing category in a supervised manner from a labeled dataset. However, their investigated body types are only two--\textit{average shape} and \textit{above average}, which is far from enough to make meaningful body shape-aware studies. In comparison, Hsiao \myet~\cite{hsiao2018creating} propose a visual body-aware embedding method that encodes fashion items with attributes and CNN features and encodes the human body with estimated SMPL~\cite{loper2015smpl} parameters and vital statistics. The item and body representation are then mapped into the joint embedding that measures body-clothing affinities. Dong \myet~\cite{dong2019personalized} determine the body shape of the users by inferring their essential body measurements from the historical clothing they have bought. They model the compatibility between the body shape and items to achieve personalized outfit recommendation.

Body shape-aware fashion recommendation is a prospective research area which is still at its early stage. Since the topic covers both the human body shape and personal preference modeling problems, both data and technical requirements are higher. From the data perspective, the collected datasets are usually on a small scale as the human body shape information is not easy to acquire. From the technical perspective, such a task may need cross-area techniques such as implicit feedback analysis in information retrieval and 3D human geometry estimation in graphics, which is a big challenge, leaving the problem under-explored so far.

\subsection{Size Recommendation}
The topic of understanding the item size and fit on a personalized level has gained momentum in the research community, especially from the industry. Research has shown that size and fit are among the most influential factors driving e-commerce customer satisfaction in the fashion domain~\cite{otieno2007fit}. Data-driven size recommendation aims to predict whether the product size is suitable for the user by analyzing the historical purchase record of the user. 
In the literature, skip-gram based method~\cite{abdulla2017size}, hierarchical Bayesian method~\cite{guigoures2018hierarchical}, graphic method~\cite{tiwari2020sizer} and etc have been proposed for size recommendation. More recently, Sheikh \myet~\cite{sheikh2019deep} propose a deep learning-based content-collaborative method to make the size/fit prediction; Dogani \myet~\cite{dogani2019learning} propose a Product Size Embedding (PSE) model to learn a latent representation for all the possible size variations of products and customers’ sizing preferences using solely purchase data and brand information; Karessli \myet~\cite{karessli2019sizenet} 
explore the size/fit pattern based on the sales and size-related returns data with a weakly supervised learning framework. Specifically, they propose a teacher-student framework with curriculum learning where the statistical model acts as the teacher and a CNN-based model (SizeNet) acts as the student that aims to learn size issue indicators from the fashion images. 
Another group of studies investigates the fits of clothing based on 3D human body~\cite{tiwari2020sizer,patel2020tailornet}, enabling to predict how the specific garments fit or drape on the body as a function of size.

Size recommendation problem is more challenging in practice than expected due to the lack of standard on size systems for fashion items such as garments and footwear. Size systems are variants in different regions and brands, which requires the size recommendation methods to not only model the relation between size and user but also explore the underlying complicated conversion logic in item sizing. One possible solution here is to combine the prior domain knowledge or certain rules instead of building the size recommendation system based solely on interaction data. For example, since the size measurement standard is known to be highly sensitive to brands, developing a brand-specific recommender system might be a more practical direction in real applications.

\subsection{Recommendation with Special Requirements}
As concerns specific requirements, various fashion recommendation tasks focusing on some detailed specific requirements have been explored in the literature, for example, recommending clothes based on the weather condition, or for special occasions. Unlike general fashion recommendation, which takes general requirements to make a recommendation, such as user preference or item compatibility, these tasks consider more additional conditions, aiming to make more specific and practical recommendations~\cite{liu2017weather,liu2012hi,verma2020fashionist,shen2007gonna}. 

For example, for the \textbf{weather-oriented recommendation} problem, Liu \myet~\cite{liu2017weather} propose a system that can automatically recommend the most suitable clothing from the user's personal clothing album, or intelligently suggest the best paired one with the user-specified reference clothing. Their model devices an additional module to train a function that can score the compatibility between the outfit images and weather conditions. So far, considering the weather in recommendation has not attracted much research attention. On one hand, it is hard to evaluate how suitable outfits are for specific weather, which means valid training data is difficult to acquire. On the other hand, the research significance of it is disputed since ordinary people could be able to avoid making big mistakes in dressing according to the weather.

In comparison, the recommendation for specific \textbf{scenarios}, \textbf{occasions}, \textbf{locations} or \textbf{scene} is of more applicable significance. An early work of Liu \myet~\cite{liu2012hi} studied two occasion-conditioned fashion recommendation problems: predicting the suitable outfit image given the occasion and predicting a specific fashion item to pair with a given item given the occasion.  They propose a latent Support Vector Machine (SVM)-based model treating clothing attributes as latent variables, with which to model the feature-occasion, attribute-occasion, and attribute-attribute potential. Similarly, Zhang \myet~\cite{zhang2017trip} work on the location-oriented recommendation, which seeks to recommend outfits for users for specific travel destinations. From the outfit image perspective, Kang \myet~\cite{kang2019complete} propose that \textit{Scene} in the outfit image has an impact on the appearance of the whole look, therefore should be considered when recommending fashion product for a given partial outfit. To address the task, they propose to measure both the item-outfit and item-scene compatibility with a category-aware visual attention mechanism. Similarly, Ma \myet~\cite{ma2019and} study the dressing for occasion problem by analyzing a large number of dressing images on Instagram, trying to derive the relationships between the clothing and occasions. Since the study is mostly based on clothing images, the performance is heavily dependent on the visual analysis of images such as clothing recognition. 

Existing work on the recommendation with special conditions mostly model the target conditions together with the general objective, for example, the compatibility between items or the personal preference of users. Therefore, we can summarize these methods as conditional recommenders. With similar a framework, we can take any influential factors into account when building the fashion recommender system.

\section{Input, Evaluation and Dataset for Fashion Recommendaiton}
\label{sec:dataset}
Input, evaluation, and dataset are important sections for empirical analysis in all algorithmic studies, including the study of fashion recommendation methods. 
In most papers reviewed above, the introduction of input, evaluation, and dataset can be found at the beginning of the experimental sections. Because of the variety of fashion recommendation tasks, the input and evaluation settings are diverse applied in different literature. There have been also a large number of datasets proposed. In this section, we summarize these three important experimental settings based on existing work, hoping to provide instructions for readers to help with their own fashion recommendation studies.

\subsection{Input of Fashion Recommendation Models}
Input is the data used to train the models in the machine learning context. For the fashion recommendation problem, input usually contains the user-item (U-I) interaction records or item-item (I-I) matching relationships, as well as side information of uses or items. For different sub-tasks in fashion recommendation, the data requirement could be different. Most existing PFR models are built based on U-I interaction records and visual information of fashion items. As summarized in Section~\ref{sec:personalized_rec}, PFR methods are developed with the incorporation of visual input updating.
On one hand, the visual information keeps being enhanced, from global to detail, keeping the input updated. On the other hand, the way of visual information incorporation has been upgraded from pre-defined features to joint-learning ones.

\begin{table}[]
    \centering
    \begin{tabular}{cc|cc|cc}
    \hline
    NO. &Pairwise Compatibility Method &NO. & Input & NO. & Evaluation \\ \hline
    1 &Topic modeling &1 & Image & 1 & PCR, Top-K accuracy \\ 
    2 &Metric learning-based &2 & Shallow visual feature &2 & PCR, AUC \\ 
    3 &Graph methods &3 & Deep visual feature &3 & FITB, top-1 accuracy \\ 
    \cline{1-2}
    NO.  &Outfit Modeling Method &4 & Text description &4 & FITB, AUC \\ 
    \cline{1-2}
   1 &Non-parametric &5 & Shallow textual feature &5 & FITB, Top-K accuracy \\ 
   2 &Sequence modeling& 6 & Deep textual feature &6 & OP, AUC \\ 
   3 &Graph modeling     & 7 & Category & 7 & Human evaluation \\
    4 &Attention mechanism & 8 & Attributes (Tags) & 8 & Others \\ 
    \hline
    \end{tabular}
    \caption{Different groups of pairwise and outfit compatibility modeling methods; Common settings of the input and evaluation in fashion recommendation experiments. PCR, FITB and OP are short for Pairwise Complementary Recommendation, Fill-In-The-Blank and Outfit Compatibility Prediction respectively. AUC denotes the Area Under the ROC Curve.}
    \vspace{-15pt}
    \label{tab:input_eval_list}
\end{table}

The input is more varied for CFR and FOR tasks. First, similar to in PFR, item-item matching relationship or outfit composition data is the basic input for most models, which is essential for the exploration of implicit compatibility patterns between item pairs or in the outfits. On top of that, since these two tasks focus more on fashion items, as well as item relationships, various content information of items from different modalities is widely used. Visual information is still dominant and has been flexibly incorporated into different models. For example, end-to-end models take item images directly as the input while some other models may use the prepared visual features extracted with shallow or deep neural network methods. Another widely adopted modality is text, which contains information resulting from multiple sources, such as 
descriptive sentences from retailers and review comments from consumers. Textual information can be applied just similar as visual features, with either the natural language input directly or high-level textual features pre-extracted by the language models. In general, the image of an item shows intuitively the look of the item and all its visual features, while textural descriptions may highlight some key features, including those that are not visible, such as brand or material. Other than the vision and language modalities, some high-level semantic information such as category or other attributes can also be leveraged in many studies of CFR or FOR. We summarize commonly applied input for item/outfit description in Table.~\ref{tab:input_eval_list}.

Since the input of different models is more complicated and varied in CFR and FOR tasks than in PFR, to further organize existing work in these two tasks, we list the input of representative models for the two tasks in Table~\ref{tab:pairwise_comp} and Table~\ref{tab:outfit_comp} respectively in Appendix. From the summary we observe that existing work in complementary/outfit recommendation can be categorized into two groups based on the input data: \textbf{uni-modal models} and \textbf{multi-modal models} by considering one or multiple modalities information as input. One particular problem in models with multi-modal input is the fusion of multi-modal content features, which can be grouped into two types: 1) pre-score and post-score fusions. Pre-score fusion methods try to obtain multi-modal features as comprehensive description of items first and then apply general projection layer for interaction/compatibility prediction~\cite{wang2019outfit, cui2019dressing,han2017learning,lin2020outfitnet}.
In comparison, post-score methods measure interaction/compatibility scores with regard to one modality and then combine the modality-specific scores together to obtain the overall multi-modal scores~\cite{song2017neurostylist,song2019gp,yang2020learning}. Laenen \myet~\cite{laenen2020comparative} conducted comparative empirical studies to specifically explore the multi-modal feature fusion problem. Their results suggest that multi-modal item representation might be more effective in outfit recommendation than uni-modal. Moreover, more sophisticated information fusions, e.g., the attention-based fusion methods, generate better item representation for outfit recommendation and bring more interpretability to the recommendation results as a by-product. However, this study focuses only on the vision-text fusion, while the fusion of more modalities or other forms of input is still not involved.

\subsection{Fashion Recommendation Evaluation}
Evaluation is an important perspective to understand a research area. Due to the specificness and variability of fashion recommendation, we find a large number of different evaluation settings in existing work, especially for complementary/outfit recommendation. After a careful survey, we systematically organize different evaluations of this research area from three perspectives: evaluation task, metric and ground truth.

\begin{figure}
    \centering
    \includegraphics[width=0.9\textwidth]{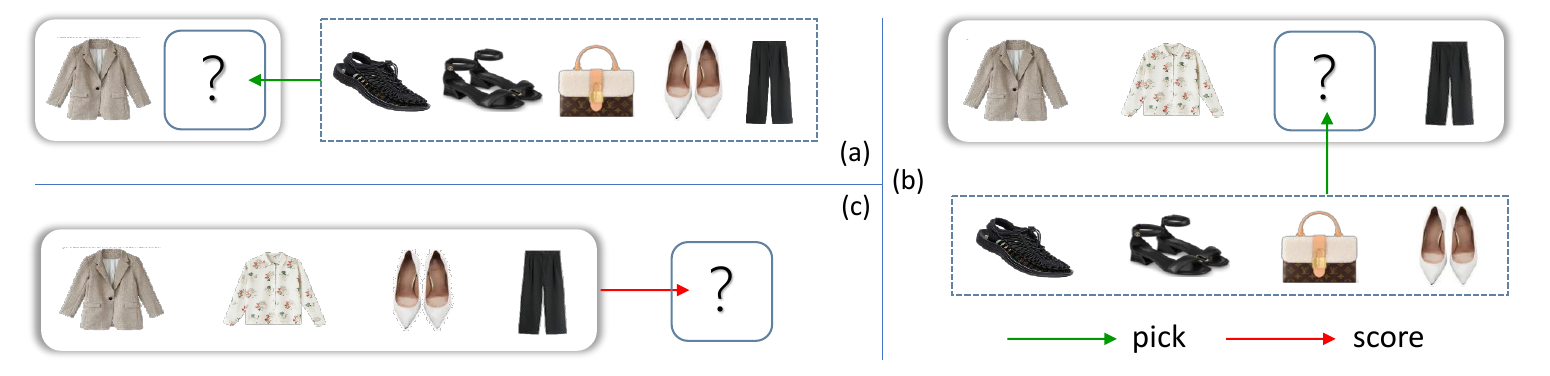}
    \caption{Three common evaluation tasks in fashion recommendation: (a) Pairwise Complementary Recommendation (PCR), (b) Fill-In-The-Blank (FITB), and (c) Outfit compatibility Prediction (OP) }
    \vspace{-12pt}
    \label{fig:comp_eval}
\end{figure}
\subsubsection{Evaluation Task}
Personalized fashion recommendation, including product recommendation (Section~\ref{sec:personalized_rec}) and outfit recommendation (Section~\ref{sec:outfit_rec}) is usually evaluated with the personalized ranking task. Specifically, given a user, the target is to predict an item for him/her, which is exactly the same as in the personalized recommendation in the general domain. In comparison, there are more evaluation tasks for complementary/outfit recommendation (Section~\ref{sec:compatibility_model} and \ref{sec:outfit_rec}). As shown in Fig.~\ref{fig:comp_eval}, the most common evaluations in literature are (a) pairwise complementary recommendation (PCR), (b) Fill-In-The-Blank (FITB)~\cite{han2017learning}, and (c) Outfit compatibility Prediction (OP). PCR requires predicting the most compatible item(s) for a given item (Fig.~\ref{fig:comp_eval} (a)) while the other two are for the evaluation of outfit compatibility modeling. Specifically, FITB needs to predict the right item from multiple choices (usually four) to complete the outfit with one item missing (Fig.~\ref{fig:comp_eval} (b)). The right item means the one that is most compatible with the given items in the outfit. This is a practical scenario in real life when people want to pick a single item to complete their look. Another task, the OP, aims to score a given set of fashion items according to the overall compatibility level to tell if they make a good outfit (Fig.~\ref{fig:comp_eval} (c)).

\subsubsection{Evaluation Metric}
Regardless of evaluation settings, there are two types of evaluation in terms of metrics calculation: pairwise comparison and Top-K ranking. The most widely used metric for the pairwise comparison setting is the Area Under the ROC Curve (AUC)~\cite{bradley1997use, rendle2010pairwise}. ROC is short for Receiver Operating Characteristic, and its curve demonstrates the accuracy of the tested learning model in predicting the positive target from two given candidates. With AUC as the evaluation metric, the testing data needs to be prepared with pairwise samples, each consisting of a positive and negative candidate. This metric is most commonly combined with the OP evaluation task to be applied in outfit compatibility modeling studies.

Top-K ranking evaluation is very commonly applied in recommendation tasks. The evaluated model first predicts a list of items (the length of the list is K) to be recommended. Then, different ranking metrics can be used to evaluate the quality of the recommended list from multiple perspectives. For example, Recall@K results demonstrate the portion of positive items that get recommended among all positive items while Precision@K results show what portion of recommended items are positive ones. Other widely used metrics include MRR@K and NDCG@K. This evaluation way is most commonly used in personalized fashion recommendation, and also in a few complementary recommendation studies. The FITB evaluation with accuracy is a special type of Top-K evaluation, which calculate the Top-1 accuracy of the item prediction for the blank. 

\subsubsection{Evaluation Ground-Truth}
In most situations, researchers save a part of the collected data for the use of evaluation, which is usually labeled properly according to the tasks. For example, for the FITB task, the test set consists of pre-defined outfits, which are all deemed as compatible. For each test outfit, one item would be randomly selected as the prediction target (positive) and removed, leaving the blank to get filled by the recommendation model to test. In this case, the ground truth is available for all test data. However, such an evaluation protocol requires additional labeling efforts, which is not applicable when lacking such labels. Meanwhile, such ground truth could be biased. For example, the composed outfit in the test set could be built by a random user in the fashion community website (\myeg~Polyvore), which means the ground truth is just the opinion of one person and can hardly perform a fair evaluation. 

To address these limitations, human evaluation has been applied. For example, to evaluate personalized recommendation models, online experiments can be conducted. Providing recommendations for users during their online activities in e-fashion platforms and recording the click-through rate (CTR) can effectively evaluate the quality of the personalized recommendations~\cite{chen2019pog} in a real-time manner. Moreover, to better evaluate compatibility learning-involved models, such as complementary recommendation or outfit composition models, some studies invite fashion experts/professionals to manually score the item pairs/outfit recommended by models regarding the compatibility~\cite{zhou2019fashion, tangseng2020toward}, making the evaluation results more sound. There is no doubt that human evaluation will perform more robust and comprehensive results. However, it is restricted by the labour cost and cannot applied in large-scale, which degrades the robustness and comprehensiveness of the evaluation results instead. Evaluation protocols applied in mainstream complementary and outfit recommendation research are also summarized in Appendix. 

\subsection{Fashion Recommendation Dataset}
To support fashion recommendation research, a large number of datasets have been constructed based on various data sources for various tasks. E-commerce platforms and online fashion communities are the two main data sources for fashion recommendation research. E-commerce data records real user--item interactions, which naturally support the research of personalized fashion recommendation. Meanwhile, the data from fashion community websites, including social media, provides more domain-specific information such as outfit composition and style evolution, which can be applied for complementary or outfit recommendation research. Due to the page limit, we summarize available datasets in Sec.~\ref{sec:dataset} in Appendix, where readers can find more details about fashion recommendation datasets.

\section{Open Challenges and Future Research Directions}
\label{sec:future_direction}
Whilst the existing research has established a solid foundation for fashion recommendation, this section outlines several promising prospective research directions. We also elaborate on several open issues, which we believe are critical to advance the present state of this field.

\subsection{Bringing More Explanation}
Explanability is a crucial yet challenging aspect of recommender systems that greatly affects the persuasiveness and user satisfaction of recommendation~\cite{vaccaro2018designing}. Currently, most fashion recommendation models cannot explain why certain items are appealing to the user, why two items match well, or why one outfit is good while the others are not. Several attempts have been made to enable more interpretability. For example, to explain the reasons for matching, Yang et al.~\cite{yang2019interpretable} propose to learn the interpretable matching patterns that lead to a good match with regard to specific attributes. Hou et al.~\cite{hou2019explainable} take advantage of semantic attributes to build an explainable fashion recommender system for personalized item recommendation. They link the user preference on items with specific attributes using a fine-grained preferences attention module. Efforts have also been made to explain the recommendation results with textual comments or descriptions~\cite{song2017neurostylist, lin2018explainable}. 

The explanations or interpretations in recommendation are mostly from two perspectives, explaining the process in the models or generating an additional explanation for the recommendation results~\cite{xiao2017attentional,seo2017interpretable,tay2018latent}. As we can see, in the specific fashion recommendation tasks, whether we analyse the process or the results, the explanations are mostly related to the attributes of items. Exploiting rich attribute information and linking it with the recommendation results is no doubt a natural and promising direction in fashion recommendation considering that user preference and matching patterns in fashion are highly related to detailed fashion attributes. In light of the particular importance of visual information in this domain, several methods focus on exploring visually the explanation for the recommendation~\cite{hou2019explainable,chen2018visually,chen2019personalized,wu2020visual}. However, the existing work only focuses on the item perspective. We should investigate further whether the patterns of users' behaviors or the connection patterns of items can be interpreted to provide an explanation.     

From the technology perspective, the attention mechanism and knowledge graph (KG) are two powerful tools to create explanations. The attention weights usually give insights about the inner workings of the models by probing the weight and activation~\cite{zhang2019deep}. This offers a more intuitive understanding of the models~\cite{xiao2017attentional,hou2019explainable}. The knowledge graph, on the other hand, empowers the model with explanability by leveraging affiliated information and jointly learning the KG and recommendation task. Such a scheme can bridge the recommendation results with specific aspects in different relations, thereby offering an explanation. For example, with an attentive attribute--aware knowledge graph, Zhan et al.~\cite{zhan2021a3} are able to characterize the user's fine-grained interests with regard to the attributes of the outfits, giving a clear explanation of the user's preference on outfits. However, although KG has been effectively applied, it is still limited in leveraging additional affiliated information and enhancing representation learning. More advanced technologies, such as machine reasoning, have been successfully applied in relevant tasks to improve explainability~\cite{zhang2019multi,tay2018multi}. Combined with knowledge graphs, machine reasoning could be the next frontier for fashion recommendation in terms of not only improving the explainability, but also achieving higher-level interactive or conversational recommendation tasks~\cite{lei2020interactive}. Causal explanation learning, which has been successfully applied in the recommendation to provide a post-hoc explanation for the black-box sequential recommendation mechanisms via causal analysis~\cite{xu2020learning}, can be another prospective way to achieve explainable recommendation.   

\subsection{Exploring and Incorporating Domain Knowledge}
The mainstream methods for fashion recommendation fit the observed data and make predictions accordingly, which overlook the rich domain knowledge. On the one hand, these methods suffer from poor interpretability as they explore the pattern in the data completely implicitly. On the other hand, ignoring additional knowledge limits the recommendation accuracy. The domain knowledge discussed here involves not only the affiliated information but also expert knowledge or rules. The latter widely exists in the fashion domain but is barely considered in previous studies. For example, traditionally, fashion matching is a human-centered task that relies on the expertise and experience of fashion stylists~\cite{dahunsi2021understanding}. In general, fashion stylists will evaluate the compatibility of two fashion items (or an outfit) from multiple dimensions, such as color or texture. They might also think about the overall styles, for example, if hip-hop ripped jeans go well with a feminine blouse. They will also consider the user perspective with respect to their body shape, color, lifestyle, etc.~\cite{dahunsi2021understanding}. Although such evaluation is subjective to some extent, there are many basic or general rules that are available on many platforms such as fashion magazines, blogs, or even YouTube videos. If this domain knowledge is effectively extracted and incorporated into the approaches developed for specific tasks, it can supplement the completely data-driven approaches and compensate for their potential deficiencies. For example, when the available data is insufficient or biased, the learned model will lack generalization and perform poorly in testing. Moreover, exploring fashion knowledge, such as matching rules, can also enhance the explanation and reliability of the recommendation results.

There are two challenges to bringing fashion domain knowledge into recommendation: 1) how to construct such knowledge; and 2) how to leverage it in different models. For the first challenge, since the knowledge is usually conveyed implicitly via different types of media, extracting structural and applicable knowledge requires extensive technologies in multi-media analysis. For the second challenge, since leveraging additional information can inevitably affect the training of the data-driven approaches, it is necessary to ensure that the incorporated knowledge is able to refine the data-fitting results rather than working completely against it and degrading the performance. Among the existing work, Ma et al.~\cite{ma2019and} focus on the fashion knowledge extraction problem and work it based on social media data. Their work explores the occasion, person, and clothing discovery from multiple modalities of images, texts, and metadata with a multi-task learning framework. Song et al.~\cite{song2018neural} for the first time incorporate fashion domain knowledge in clothing matching. Specifically, they present an attentive knowledge distillation scheme to encode the knowledge to the traditional neural networks to boost the compatibility modeling performance. Apart from these attempts, the relevant studies are still limited. On the one hand, many data sources are available and valuable for fashion knowledge extraction but have not been focused on yet. On the other hand, more advanced technologies need to be developed to extract the best of the available knowledge to facilitate specific fashion recommendation tasks.

\subsection{Modeling Heterogeneous Content Information and Multiple Behaviors}
Fashion recommendation is a task that needs to process heterogeneous data, including implicit feedback records, user profiles, item attributes, review text, item combination relationships, and other auxiliary information. However, most of the existing methods fail to utilize the various types of side information in a comprehensive manner and take full advantage of the available data. Many of them build the recommender system relying on the implicit feedback or the observed item combination data, mostly with visual information. Considering the exciting recent advances in graph neural networks (GNNs), modeling the heterogeneous data in fashion recommendation tasks with heterogeneous graphs~\cite{wang2019heterogeneous, zhang2019heterogeneous} is a promising direction. In a heterogeneous graph, different types of nodes (users, items, attributes, etc) can be related in different ways (interacting, affiliated, etc.). GNNs have been applied in several tasks, such as compatibility modeling and outfit recommendation. In compatibility modeling, several methods use a graph to model multiple relations between fashion items to enhance the item representation by conducting information propagation and aggregation on the graphs~\cite{yang2020learning,cucurull2019context}. For outfit recommendation, the graph is usually used to encode the outfit with several items and to obtain the outfit representation~\cite{lin2020outfitnet,li2020hierarchical,cui2019dressing}. In addition, the knowledge graph has been successfully applied to incorporate attributes and facilitate the outfit preference prediction~\cite{zhan2021a3}. 

However, the use of graph technologies in fashion recommendation is still under-explored. 
The first challenge here is how to design such a graph for different specific tasks that can effectively incorporate as much useful information as possible. Furthermore, how to jointly model the abundant auxiliary information to facilitate the recommendation task, for example, the user preference or item compatibility modeling, is another challenging aspect. To address these challenges, effective heterogeneous graph learning or multi-task learning approaches should be developed.

\subsection{Solid Benchmark and Evaluation}
In the previous sections, we have mentioned that both the compatibility modeling and outfit recommendation tasks lack high-quality benchmarks. In various studies, tens of datasets are used for the same task, which makes it hard to fairly evaluate different methods and analyse the development of the research field. To facilitate the research in the mentioned tasks, standard task settings, as well as an evaluation dataset and metrics should be developed. For compatibility modeling, most of the existing work is based on Polyvore datasets but uses different versions of data with different statistics. In fact, one or several benchmark datasets can be developed based on the existing Polyvore datasets to at least fix the evaluation part of the data. For the outfit recommendation task, both the datasets and the task need a standard, for example, whether the outfits are pre-defined. 

On top of the need to improve the benchmarks, current evaluation schemes need improvement too. For personalized fashion recommendation, the existing work generally applies a small negative set evaluation. However, there have been studies arguing that small negative set evaluations can lead to unfair evaluation results~\cite{krichene2020sampled}. As a result, in the general domain, full rank evaluation has been widely applied in recent years~\cite{wang2019ngcf,he2020lightgcn}. However, full rank evaluation is in some way not applicable to the fashion domain as the item set is large-scale (tens of thousands or even millions). Applying full-rank evaluation will require substantial computer power and may not even be applicable. More importantly, the values of the generally used metrics such as recall@10 or ndcg@10 will be too small to effectively evaluate the methods. Therefore, a proper evaluation scheme needs to be designed for this task. 

\subsection{Understanding Better the Needs of Fashion Industry and Users}
The ultimate goal of fashion recommendation technology is to help the fashion industry to make wise business strategies, and users to make good purchasing or dressing decisions. Therefore, for the benefit of fashion recommendation research, on the one hand, we should enhance the benchmark and standard setting for basic tasks to support the development of fundamental techniques, as we state above. On the other hand, research on fashion recommendation should catch up with the development of the fashion industry, being aware of the real needs of the industry and where it is heading.

From the business perspective, the basic idea of current fashion recommendation is to attract customers by offering the item options they might like. Such a strategy is passive to some extent. Traditional fashion marketing strategies such as promotion and advertisement can effectively facilitate the retailing business, positively creating demand from consumers rather than capturing the existing ones~\cite{easey2009fashion}. However, these traditional marketing strategies are usually addressed to certain customer groups, and not personalized. Therefore, the first potential direction would be to develop recommendation algorithms with marketing functions, making them able to effectively make promotions to customers like the traditional fashion marketing strategies but to customize and personalize these promotions. This is a challenging problem that needs a deep understanding of fashion marketing knowledge and to further incorporate certain theories and strategies with the data-driven recommender systems. 

Another big goal is omnichannel fashion recommendation.  An omnichannel strategy seamlessly integrates the different sales and communication channels that the customer uses, combining the strengths of each channel~\cite{lorenzo2020omnichannel, lynch2020omnichannel,mcintyre2016multi}. To facilitate omnichannel retail in the technical aspect, we need to collect data from multiple channels and build data-driven recommendation models, which is a great challenge, for example, needing to extract, align and make use of user and product information across channels. Most previous studies on fashion recommendation are limited to one data channel, for example, applying e-commerce data for personalized recommendation or social media data for the mix-and-match recommendation. In other domains of recommendation, researchers have proposed learning a general user representation for multiple recommendations across platforms~\cite{yuan2021one, yuan2020parameter} with the transfer learning~\cite{bertinetto2016learning} and continual learning~\cite{li2017learning,zenke2017continual} technologies. This represents a promising direction for us to develop omnichannel fashion recommendation. Besides, there might be more technical parts required to realize the overall omnichannel recommendation. For example, how can we capture the latest fashion trends from social media~\cite{ding2022leveraging} or the dressing knowledge~\cite{ma2019and}, and then leverage them for fashion recommendation on e-commerce platforms. 

From another perspective, users' needs for fashion recommendation are not limited to the purchase processes, or we can say that nowadays buying new clothing is no longer the only thing that people want in terms of fashion. One report shows that Gen-Z and Millennial consumers are more eco-conscious and have a growing interest in second-hand clothes\footnote{https://www.businessoffashion.com/articles/retail/affirm-evolution-omnichannel-retail}. Such an area of second-hand fashion recommendation has barely been explored in the literature so far, yet has great applicable values. We can expect that such a task will have to deal with a severe sparsity and long-tail issue~\cite{yin2012challenging}, which can make most CF-based recommendation methods fail and requires a better understanding of products semantically~\cite{he2016vista}. Another real need from ordinary users is to make the best use of their own wardrobe, which is mostly related to the existing task of, capsule wardrobe creation (see section~\ref{sec:outfit_rec} for details). Furthermore, with the increasing influence of minimalism, how to optimize the personal wardrobe by substituting minimal items is another practical problem that ordinary users care about. All in all, the development of fashion recommendation should be closely integrated with the practical needs from both the business and user perspectives.

\section{Conclusions}
\label{sec:conclusion}
This paper focused on the field of fashion recommendation and extensively reviewed the recent research efforts in this field. We proposed a classification scheme for organizing existing works based on the task settings, which divides all fashion recommendation studies into personalized fashion recommendation, complementary recommendation, outfit recommendation and special fashion recommendation. We comprehensively summarized the existing methods for each group of works and pointed out the limitations of the current research. By reviewing all these sub-tasks in fashion recommendation, we outlined the development of this research field. We further discussed several promising directions in fashion recommendation for future extension from different perspectives, including narrowing the gap between the real needs of the industry and the research outcomes. Overall, this survey provided a comprehensive and systematic review of the existing literature on fashion recommendation, and also offered a deep insight into how the fashion industry could benefit further from these academic developments.

\bibliographystyle{ACM-Reference-Format}
\bibliography{10.bib-refs}
\newpage
\appendix
\section{Appendix}
\setcounter{table}{0}
\setcounter{footnote}{0}
\renewcommand\thetable{\Alph{section}\Roman{table}} 
\begin{table}[h]
    \centering
    \begin{tabular}{p{3mm}p{70mm}p{8mm}p{7mm}p{8mm}p{7mm}p{9mm}p{14mm}}
        \hline
         NO. & Paper & Where & When & Method &Input & Dataset & Evaluation \\ \hline
         1 & Fashion Coordinates Recommender System Using Photographs from Fashion Magazines \cite{iwata2011fashion} 
         & IJCAI  & 2011 &1 & 2 & 18 & 1 \\
         2 & Large Scale Visual Recommendations From Street Fashion Images \cite{jagadeesh2014large} 
         & KDD    & 2014 &1 &3 &20 &7\\
         3 & Learning Visual Clothing Style with Heterogeneous Dyadic Co-occurrences \cite{veit2015learning} 
         & ICCV   & 2015 &2 &1 &29 &2,7 \\
         4 & Learning Compatibility Across Categories for Heterogeneous Item Recommendation \cite{he2016learning} 
         & ICDM   &2016  &2 &3 &29 & 8 \\
         5 & Neurostylist: Neural Compatibility modeling for Clothing Matching \cite{song2017neurostylist} 
         & MM     & 2017 &2 &3,6 &8,10 &1,2 \\
         6 & Compatibility Family Learning for Item Recommendation and Generation \cite{shih2018compatibility} 
         & AAAI   &2018 &2 &3 &2,29 &2 \\
         7 & Neural Compatibility Modeling with Attentive Knowledge Distillation \cite{song2018neural} 
         & MM     &2018 &2 &3,6 &3 &2 \\
         8 & Fashion Sensitive Clothing Recommendation Using Hierarchical Collocation Model \cite{zhou2018fashion} 
         & MM     &2018 &2 &2 &32,33 &1\\
         9 & Learning Type-aware Embeddings for Fashion Compatibility \cite{vasileva2018learning} 
         & ECCV   & 2018 &2 &3,6 &6,7 &3,6\\
         10 & Learning to Match on Graph for Fashion \cite{yang2020learning} 
         & AAAI   & 2019 &3 &3,6 & 29 &1\\
         11 & TransNFCM: Translation-Based Neural Fashion Compatibility Modeling \cite{yang2019transnfcm} 
         & AAAI   & 2019 &3 &3,6 &2,3 &1, 2 \\
         12 & Interpretable Fashion Matching with Rich Attributes \cite{yang2019interpretable} 
         & SIGIR  & 2019 &/ &3,6,8 &24 &1\\
         13 & Context-aware Visual Compatibility Prediction \cite{cucurull2019context} 
         & CVPR   & 2019 &3 &3 &2,29 &3,6\\
         14 & Learning Similarity Conditions Without Explicit Supervision \cite{tan2019learning} 
         & ICCV   & 2019 &2 &3,6 &2,7 &3,6 \\
         15 & Improving Outfit Recommendation with Co-supervision of Fashion Generation \cite{lin2019improving} 
         & WWW    & 2019 &2 &1,4 &3,8 &1,2\\
         16 & Explainable Outfit Recommendation with Joint Outfit Matching and Comment Generation \cite{lin2018explainable} 
         & TKDE   & 2019 &/ &1,4 &3,8 &1,8\\
         17 & Toward Explainable Fashion Recommendation \cite{tangseng2020toward}
         & WACV   & 2020 &/ &2 &5 &7\\
         18 & Fashion Compatibility Modeling through a Multi-modal Try-on-guided Scheme \cite{dong2020fashion} 
         & SIGIR  & 2020 &/ &3 &41 &1,6\\ 
         19 & Fashion Outfit Complementary Item Retrieval~\cite{lin2020fashion} &CVPR &2020 &2 &1,7 &7 &3,6 \\
         \hline
    \end{tabular}
    \caption{Summary of mainstream research on complementary fashion recommendation. Methods, input and evaluation settings refer to Table.~\ref{tab:input_eval_list}, and `/' denotes \textit{other methods} that are not listed in the table. Dataset refers to Table.~\ref{tab:dataset}. }
    \label{tab:pairwise_comp}
\end{table}
\newpage

\begin{table}[]
    \centering
    \begin{tabular}{p{3mm}p{73mm}p{11mm}p{7mm}p{8mm}p{6mm}p{8mm}p{12mm}}
    \hline
    NO. & Paper & Where & When & Method &Input & Dataset & Evaluation \\ \hline
    1 & Recommending Outfits from Personal Closet \cite{tangseng2017recommending} & ICCV(W) & 2017 &1 &3 &5 &8 \\
    2 & Learning Fashion Compatibility with Bidirectional LSTMs \cite{han2017learning} & MM & 2017 &2 &3 &2 &3,6\\
    3 & Mining Fashion Outfit Composition Using An End-to-End Deep Learning Approach on Set Data \cite{li2017mining} & TMM & 2017 &2 &3,5,7 &4 &6 \\
    4 & Interpretable Partitioned Embedding for Customized Multi-item Fashion Outfit Composition \cite{feng2018interpretable} & ICMR & 2018 &3 &2 &13 &7 \\
    5 & Outfit Recommendation with Deep Sequence Learning \cite{jiang2018outfit} & BigMM & 2018 &2 &1,4 &2 &8\\

    6 & Outfit Compatibility Prediction and Diagnosis with Multi-Layered Comparison Network \cite{wang2019outfit} & MM & 2019 &/ &1,4 & 12 &3,6\\
    7 & Dressing as a whole: Outfit Compatibility Learning Based on Node-wise Graph Neural Networks \cite{cui2019dressing} & WWW & 2019 &3 &3,6 &2 &3,6\\
    8 & Low-rank Regularized Multi-representation Learning for Fashion Compatibility Prediction \cite{jing2019low} & TMM & 2019 &/ &2 &2 &8\\
    
    9 & Towards a Unified Framework for Visual Compatibility Prediction \cite{singhal2020towards} & WACV & 2020 &3 &3,7 &2,7 &3\\
    10 & Learning Tuple Compatibility for Conditional Outfit Recommendation \cite{yang2020learning2} & MM & 2020 &4 &3,7 &40 & 3,6\\ 
    11 & Learning Diverse Fashion Collocation by Neural Graph Filtering~\cite{liu2020learning} & TMM & 2020 &3 &3 &2,6,31 &3,6\\ 
    12 & Learning the Composition Visual Coherence for Complementary Recommendation~\cite{li2020learning} &IJCAI &2021 &4 &3 &2,7 &3,5\\
    \hline
    \end{tabular}
    \caption{Summary of mainstream research on outfit compatibility modeling. Methods, input and evaluation settings refer to Table.~\ref{tab:input_eval_list}, and `/' denotes \textit{other methods} that are not listed in the table. Dataset refers to Table.~\ref{tab:dataset}. }
    \label{tab:outfit_comp}
\end{table}
\newpage

\subsection{Dataset Summary}
\label{sec:dataset}
\subsubsection{Data Sources}
E-commerce platforms and online fashion communities are the two main data sources for fashion recommendation research. E-commerce data records real user--item interactions, which naturally support the research of PFR. Meanwhile, the data from fashion community websites, including social media, provides more domain-specific information such as outfit composition and style evolution, which can be applied for CFR or OR research. So far, most existing fashion recommendation datasets were developed from the following sources:

\begin{itemize}
    \item \textbf{Polyvore}\footnote{https://en.wikipedia.org/wiki/Polyvore} is a fashion website that enables fashion lovers to create outfits as compositions of clothing items. Polyvore provides a large number of high-quality fashion outfits. Moreover, each item for the composition of outfit is associated with rich content information including the image, descriptive tags and text, category information etc., making it a great data source for outfit-related studies such as compatibility modeling, outfit generation and recommendation. To this end, many fashion datasets have been constructed in previous studies based on Polyvore.
    
    \item \textbf{Taobao}\footnote{https://www.taobao.com} is a large-scale online-shopping platform, which is also the largest online fashion shopping platform in China. Fashion is one of the biggest and the oldest businesses of Taobao. A new application iFashion has been created to support fashion outfit recommendation. Approximately 1.5 million content creators were actively supporting Taobao as of March 31, 2018.
    
    \item \textbf{Amazon}\footnote{https://www.amazon.com} is one of the largest world-wide e-commerce platforms based in the US and started its business as bookseller. For now, fashion has become one of the main product categories sold on Amazon. Both the Taobao and Amazon provides real-world user-involved interaction records in online shopping context, which are of great value to develop models for practical use. 
    
    \item \textbf{Other} fashion communities sources include Lookbook, Chictopia and Instagram\footnote{https://lookbook.nu/, https://chictopia.nu/, https://www.instagram.com/}. Other e-commerce sources include eBay, ASOS, JD\footnote{https://www.ebay.com/, https://www.asos.com/, https://www.jd.com} and some single brand websites. 
\end{itemize}

\subsubsection{Existing Datasets}
We present the organization of the existing datasets in Table~\ref{tab:dataset}, which contains the publication year, the basic statistics, and a brief introduction to each dataset. From the dataset sorting, we can see that more than 30 datasets are proposed in the literature for fashion recommendation tasks, which are quite rich in kind. Although these datasets might be for different sub-tasks, the data options for some specific tasks are still too many. For example, based on Polyvore, around 14 datasets have been constructed, most of them applied for compatibility modeling or outfit recommendation. However, there is still no benchmark for certain tasks to fairly evaluate different approaches. Most existing datasets have been applied only once in evaluating the corresponding methods, which is sometimes unnecessary. Even though some methods want to emphasize certain features or information in addressing certain tasks, they could have expanded or improved the existing datasets rather than proposing a new one, usually introducing too many differences in the new dataset. For example, almost every Polyvore-based dataset has side information on the item category, but such information can be different, being fine-grained in some datasets (Li's Polyvore categorizes items into 333 categories) but coarse in others (only five categories). Such differences in details can also affect the evaluation of different methods. 

We can conclude that there is so far no satisfactory benchmark dataset for any sub-task in fashion recommendation. For personalized fashion recommendation, most methods have been developed based on the Amazon Review dataset~\cite{kang2017visually, mcauley2015image} (No. 23 in Table~\ref{tab:dataset}), and are therefore more consistent in their development and more comparable. However, only applying one dataset somehow makes the results biased and unable to comprehensively evaluate the methods. Even though it is hard to find another dataset so far that has been specifically proposed for personalized fashion recommendation, we can create one based on the available extensive datasets. The iFashion dataset (No. 26 in Table~\ref{tab:dataset}) was originally for the outfit recommendation task. But it provides the sequences of the user's successive clicking on items, offering real-world user-item interaction data with another type of behavior and from another source, which can serve as a great complement for the Amazon dataset (No. 23) for the personalized fashion recommendation task. 
 
For compatibility learning or complementary recommendation, as we mentioned above, although the available datasets are many, few of them have been applied more than once. In fact, most datasets are from three main sources: Polyvore, Taobao, and IQON, which have different characteristics, such as different outfit styles (western, Chinese, and Japanese). By combining the three sources of data, more solid benchmark datasets can be developed.

\newpage
\begin{center}
\setlength{\tabcolsep}{4.5pt}
\begin{longtable}[t]{|p{0.02\textwidth}<{\centering}|p{0.15\textwidth}<{\centering}|p{0.04\textwidth}<{\centering}|p{0.04\textwidth}<{\centering}|p{0.04\textwidth}<{\centering}|p{0.05\textwidth}<{\centering}|p{0.05\textwidth}<{\centering}|p{0.33\textwidth}<{\centering}|p{0.12\textwidth}<{\centering}|}
    \caption{Existing datasets for fashion recommendation grouped by data sources. The numbers and quantities are reported approximately. K denotes thousands and M denotes millions. Abbreviation is used to denote different item information as well, C for Category, T for Textual description, including titles, I for Image, V for Votes, A for Attributes, O for Occasion, L for Location, S for Style, and H for Hashtag or other tags. NA denotes not applicable and / denotes the information is not given in the original papers. }
    \label{tab:dataset} \\ 
    \hline
    No.  & Dataset  &Year &\#User &\#Item &\#Outfit & Item Info &Description &Applied in  \\ \hline
    \multicolumn{9}{|c|}{Polyvore} \\ \hline
        1 & "Sets" on Polyvore~\cite{hu2015collaborative} & 2015 & 150 &83K &  NA
        &C
        &Outfit = Top + Bottom + Shoes. 
        &\cite{hu2015collaborative} \\ \hline
        
        2 & Maryland Polyvore~\cite{han2017learning} & 2017 & NA & 16K & 22K 
        &C, T
        &Outfit lengths are varied and less than 8.
        & \cite{han2017learning,cui2019dressing,jiang2018outfit,jing2019low,nakamura2018outfit,cucurull2019context,singhal2020towards,yang2019transnfcm,li2020learning,tan2019learning}\\ \hline
         
        3 & FashionVC \cite{song2017neurostylist} & 2017 & 248 & 29K & 21K pairs 
        &C, I, T 
        &Outfit = Top + Bottom.
        &\cite{song2017neurostylist,yang2019transnfcm,li2020fashion,lin2019improving,lin2018explainable} \\ \hline

        4 & Li's Polyvore \cite{li2017mining} & 2017 & NA  & 368K & 195K 
        &C, T
        & Outfit length is 4; Category taxonomy is unstructured.
        &\cite{li2017mining}\\ \hline
        
        5 &Polyvore 409~\cite{tangseng2017recommending} &2017 & NA &644K &409K 
        &C, I, T
        &Outfit lengths are varied; Likes and comments of outfits are given.
        &\cite{tangseng2017recommending, tangseng2020toward} \\ \hline
        
        6 & Polyvore Outfits-D~\cite{vasileva2018learning} & 2018 & NA  & 32K & 175K 
        &C, T
        &Outfit lengths are varied and less than 16; categories are fine-grained.
        &\cite{vasileva2018learning}\\ \hline
        
        7 & Polyvore Outfits~\cite{vasileva2018learning} & 2018 & NA  & 365K & 68K
        &C, T  
        &Outfit lengths are varied and less than 19; categories are fine-grained.
        &\cite{vasileva2018learning,cucurull2019context,li2019coherent,li2020learning,liu2020learning,singhal2020towards,tan2019learning,kim2020self,lin2020fashion}\\ \hline
        
        8 &ExpFashion \cite{lin2018explainable}  & 2018 & NA & 51K &201K 
        &C
        &Outfit = Top + Bottom; Each outfit has several comments.
        &\cite{lin2018explainable}\\ \hline
        
        9 &Capsule \cite{hsiao2018creating} &2018 &NA &7,478 &3,759 &C, I, H &Each item is labeled with the season, occasion and function to wear it. &  \cite{hsiao2018creating,kim2020self}\\ \hline
        
        10 & FashionVC+ \cite{liu2019neural} & 2019 & 248 & 29K & 21K 
        &C, I, T
        &Outfit = Top + Bottom + Shoes 
        &\cite{liu2019neural} \\ \hline
        
        11 &Polyvore-U~\cite{lu2019learning} & 2019 & 630 & 205K & 150K 
        &C, I
        &Outfit = Top + Bottom + Shoes; Four versions of data in total and the largest version is shown here.
        &\cite{lu2019learning,lu2021personalized}\\ \hline
        
        12 &Polyvore-T~\cite{wang2019outfit} & 2019 & NA  & 84K & 20K 
        &C, I, T, V
        &Each outfit contains 3-5 items belonging to main categories. 
        &\cite{wang2019outfit}\\ \hline
        
        13 & Feng's Polyvore~\cite{feng2019interpretable}  & 2019 & NA  & 7.53M & 1.56M 
        &C, I, V
        &Each outfit is associated with the number of likes. 
        &\cite{feng2019interpretable}\\ \hline

        14 &Polania's polyvore~\cite{polania2019learning} &2019 & NA & NA &14K 
        & /
        &Each outfit is composed of at least 2 items. 
        &\cite{polania2019learning}\\ \hline
        
        15 & Lin's polyvore~\cite{lin2020outfitnet}  & 2020 & 150 & 159K & 66K 
        &  /
        &Over 68\% of fashion outfits are liked by one user only.  
        &\cite{lin2020outfitnet} \\ \hline
        
        16 &EVALUATION3 \cite{zou2020regularizing} &2020 &NA & NA &18K pairs 
        &I, A
        &Each top-bottom pair is labeled for its fashionability (good/bad/normal) by fashion experts with reasons. 
        &\cite{zou2020regularizing} \\ \hline
        
        17 &Moosaei's Polyvore~\cite{moosaei2020fashion} &2020 &NA &256K &50K 
        &I, T 
        &Included users have over 100K followers; Each outfit contains at least 2 items, and five on average. 
        &\cite{moosaei2020fashion}\\ \hline

    \multicolumn{9}{|c|}{Other fashion communities (Instagram, Flickr, Pinterest, chictopia, etc)} \\ \hline
        18 &Magazine \cite{iwata2011fashion} & 2011 &NA &NA &14,813 &NA & &\cite{iwata2011fashion} \\ \hline
        
        19 & WOW (Flickr)~\cite{liu2012hi} &2012  &NA & 34K & 24K 
        &A, O
        &It is specifically for occasion-oriented clothing recommendation. 
        &\cite{liu2012hi}\\ \hline
        
        20 & Fashion-136K (street)~\cite{jagadeesh2014large} & 2014 & 8357 & NA & 136K 
        &I, H, B, L
        &It contains street photos in various poses and complex backgrounds. 
        &\cite{jagadeesh2014large} \\ \hline
        
        21 &Fashion144k (Chictopia)~\cite{simo2015neuroaesthetics} &2015 & 14,287 &NA &144K &I, T, H &Each post contains an outfit, containing 3.22 fashion items on average. &\cite{simo2015neuroaesthetics}\\ \hline
        
        22 &Street Fashion (street)~\cite{zhou2018fashion} & 2018 &NA &NA & 1M pairs 
        & /
        &The fashion pairs are generated from 100K street photographs. 
        &\cite{zhou2018fashion}  \\ \hline
        
        23 &FashionKE (Instagram)~\cite{ma2019and} &2019 &NA &NA &81K 
        &C, A, O
        &It contains full-body human image.  
        &\cite{ma2019and,verma2020fashionist} \\ \hline
        
        24 & Lookastic (Lookastic)~\cite{yang2019interpretable}  &2019  &NA  &15K  &31K  
        &C, I, A
        &The dataset is divided into two parts: women and men.  
        &\cite{yang2019interpretable}\\ \hline
        
        25 &Fashion Takes Shape (Chictopia)~\cite{sattar2019fashion} &2019 &180 &NA &18K 
        &C, I
        &Each image is labeled with the body shape (average or above average) of the user. 
        &\cite{sattar2019fashion}\\ \hline
        
        26 &Shop the Look (Pinterest)~\cite{kang2019complete} &2019 &NA &38K &72K
        &C, I &Outfit = Top + Bottom; Both the product and scene-product pair images are provided; Product is labeled with bounding box in images. &\cite{kang2019complete} \\ \hline
        
        27 &Fashionist \cite{verma2020fashionist,verma2020addressing} &2020 &NA &NA &2893 
        & C, A, O
        &It contains fashion images with natural backgrounds.
        &\cite{verma2020fashionist,verma2020addressing} \\ \hline 
        
        28 &SocialMediaRec (Lookbook)~\cite{zheng2020personalized} &2020 &2293 &NA &229K 
        &C, T, H
        &Included users have over 7K fans and 100 selfie posts; Age, likes and fans of users are given. 
        &\cite{zheng2020personalized} \\ \hline
        
    \multicolumn{9}{|c|}{Amazon} \\ \hline
        29 & Clothing, Shoes and Jewelry~\cite{mcauley2015image,kang2017visually} & 2015  &NA  &1.5M (2014) 2.7M (2018)  &NA  
        &T, I, C
        &It provides detailed user-item interaction information including reviewer, rating and time; Price and brand information is partially given. 
        & \cite{mcauley2015image,kang2017visually,vbpr,he2016sherlock,yu2018aesthetic,hou2019explainable,veit2015learning,shih2018compatibility,yang2020learning,cucurull2019context,chong2020hierarchical} \\ \hline
        
        30 &bodyFashion \cite{dong2019personalized} &2019 &12K &76K &NA 
        &C, I, T
        &User-item interaction records with size and rating on items. 
        &\cite{dong2019personalized} \\ \hline
        
        31 & Amazon Fashion Dataset~\cite{liu2020learning} & 2020  &NA  &NA &60K 
        &C, O, S 
        &This dataset is specifically for fashion collocation. 
        &\cite{liu2020learning,sun2020learning} \\ \hline

    \multicolumn{9}{|c|}{Taobao} \\ \hline
        32 &Taobao~\footnote{https://tianchi.aliyun.com/dataset/dataDetail?dataId=3326} &2017 &NA &61k & 406k pairs 
        &C
        & Item pairs are matched by the fashion experts.
        &\cite{zhao2017deep,zhou2018fashion} \\ \hline
        
        33 & OSA (Taobao, JD, etc)~\cite{zhou2018fashion} & 2018 &NA  & 20k & NA
        &  /
        & Each item is described with 2-3 images of different human postures in various background. 
        &\cite{zhou2018fashion}\\ \hline
        
        34 & iFashion~\cite{chen2019pog} & 2019 & 3.6M & 4.5M & 127K 
        &C, I, T
        &Each outfit has at least 4 items; Each user interacts with over 40 outfits.
        &\cite{chen2019pog,li2020hierarchical,lin2020outfitnet} \\ \hline

    \multicolumn{9}{|c|}{Other e-commerce platforms (IQON, Net-A-Porter, etc)} \\ \hline
        35 &E-commerce matching (Net-A-Porter, etc) \cite{liu2018learning} &2018 &NA &50K &68K pairs 
        &I, T
        &Images for items are multi-view; Item descriptions covers information of category, brand, price etc; Items are matched by the fashion experts. 
        &\cite{liu2018learning}\\ \hline
        
        36 & Style4Body -Shape\cite{hidayati2018dress} &2018 &270 &NA &348K &I & Body measurements of different users are given. &\cite{hidayati2018dress, hidayati2021body} \\ \hline
        
        37 &IQON (IQON)~\cite{nakamura2018outfit} &2018 &NA &200K &89K 
        & I
        &Each outfit is liked for at least 50 times.
        &\cite{nakamura2018outfit} \\ \hline
        
        38 & IQON3000 (IQON)~\cite{song2019gp} & 2019 & 3,568 & 672K &309K 
        &C, I, A, T
        &Each outfit has the price information and the number of likes. &~\cite{song2019gp,zhan2021a3,lu2021personalized,Sagar2020paibpr} \\ \hline
        
        39 & Zhou (Zalora, ASOS, etc)~\cite{zhou2019fashion} & 2019 &NA  &NA  & NA
        &C, I, T
        & It contains full-body, half-body, product and detailed images, as well as the brand and price information.   
        &\cite{zhou2019fashion}\\ \hline
        
        40 & IQON (IQON)~\cite{yang2020learning2} & 2020 & NA & NA & 29K 
        &C, I
        & Each outfit has been liked by over 70 people; Categaries are fine-grained.
        & \cite{yang2020learning2}  \\ \hline

        41 &FOTOS (SSENSE)~\cite{dong2020fashion} &2020 &NA &11K &20K 
        &I, T
        & Each outfit is worn by one model in standard front-view post. 
        &\cite{dong2020fashion}\\ \hline
        
        42 &VIBE (Birdsnest)~\cite{hsiao2020vibe} &2020 &68 &1957 & NA
        &I, T
        &It contains dresses worn by models different body shapes and the measurements are given. 
        &\cite{hsiao2020vibe}\\ \hline
        
        43 & \cite{verma2020addressing} &2020 &NA &NA &2893 &C, I, A, O & User information such as age and gender are given. &\cite{verma2020addressing, verma2020fashionist} \\ \hline
\end{longtable}
\end{center}

\end{document}